\documentclass[twocolumn,10pt]{article} 

\usepackage[square,numbers,sort&compress,comma]{natbib}

\usepackage{amsmath}
\usepackage{amssymb}
\usepackage{caption}
\usepackage{graphicx}
\usepackage{latexsym}
\usepackage{times}
\usepackage[pagewise]{lineno}
\usepackage{hyperref}

\topmargin - 12pt 
\renewenvironment{abstract}%
              {
               \small
               {\bfseries \abstractname}
               \par
               \vspace{10pt}
              }

\renewcommand\abstractname{Abstract}

\newcommand{\nomenclature}
              [1]
              {
               \bgroup
               \flushleft
               \small\bf
               #1
               \par
               \egroup
              }

\renewcommand{\section}
              [1]
              {
               \bgroup
               \flushleft
               \small\bf
               \refstepcounter{section}
               \arabic{section}. #1
               \par
               \egroup
              }

\renewcommand{\subsection}
              [1]
              {
               \bgroup
               \flushleft
               \small\em
               \refstepcounter{subsection}
               \arabic{section}.
               \arabic{subsection}. #1
               \par
               \egroup
              }

\renewcommand{\subsubsection}
              [1]
              {
               \bgroup
               \flushleft
               \small\em
               \refstepcounter{subsubsection}
               \arabic{section}.
               \arabic{subsection}.
               \arabic{subsubsection}. #1
               \par
               \egroup
              }

  \newcommand{\acknowledgement}
              [1]
              {
               \bgroup
               \flushleft
               \small\bf
               #1
               \par
               \egroup
              }

  \newcommand{\sectionbib}
              [1]
              {
               \bgroup
               \flushleft
               \small\bf
               #1
               \par
               \egroup
              }

\begin{document}

\title{\LARGE Experimental and Computational Investigation of the Influence of Ethanol on Auto-ignition of {\it n-}Heptane in Non-Premixed Flows\\
                    }

\author{{\large Liang Ji, Kalyanasundaram Seshadri$^{*}$, Forman A. Williams}\\[10pt]
        {\footnotesize \em University of California at San Diego, La Jolla, California 92093-0411, USA }}

\date{}


\small
\baselineskip 10pt


\twocolumn[\begin{@twocolumnfalse}
\vspace{50pt}
\maketitle
\vspace{40pt}
\rule{\textwidth}{0.5pt}
\begin{abstract} 
\newcommand{\eqn}{Eq.}
\newcommand{\eqns}{Eqs.}
\newcommand{\aea}{activation-energy asymptotic}
\newcommand{\caea}{Activation-energy asymptotic}
\newcommand{\caeo}{Activation-energy}
\newcommand{\aeo}{activation-energy}
\newcommand{\aeays}{activation-energy asymptotic analyses}
\newcommand{\rra}{rate-ratio asymptotic}
\newcommand{\crra}{Rate-ratio asymptotic}
\newcommand{\asyan}{asymptotic analysis}
\newcommand{\asyans}{asymptotic analyses} 
\newcommand{\ci}{chemical inhibition}
\newcommand{\chc}{characteristic}
\newcommand{\sr}{strain rate}
\newcommand{\sdr}{scalar dissipation rate}
\newcommand{\fl}{flame}
\newcommand{\cf}{counterflow}
\newcommand{\cp}{c_{\rm {p}}}
\newcommand{\cfg}{counterflowing}
\newcommand{\df}{diffusion flame}
\newcommand{\ppf}{partially-premixed}
\newcommand{\cppf}{Partially-premixed}
\newcommand{\pf}{premixed flame}
\newcommand{\pfs}{premixed flames}
\newcommand{\np}{nonpremixed}
\newcommand{\npf}{nonpremixed flame}
\newcommand{\npfs}{nonpremixed flames}
\newcommand{\pheat}{preheat zone}
\newcommand{\post}{post-flame zone}
\newcommand{\zeld}{Zel'dovich number}
\newcommand{\as}{air stream}
\newcommand{\fs}{fuel stream}
\newcommand{\frs}{fuel-rich stream}
\newcommand{\fls}{fuel-lean stream}
\newcommand{\fsd}{fuel side}
\newcommand{\frbd}{fuel-rich boundary}
\newcommand{\flbd}{fuel-lean boundary}
\newcommand{\fbd}{fuel boundary}
\newcommand{\os}{oxidizer stream}
\newcommand{\osd}{oxidizer side}
\newcommand{\obd}{oxidizer boundary}
\newcommand{\ma}{methane air} 
\newcommand{\ck}{chemical-kinetic}
\newcommand{\cbk}{chain-breaking}
\newcommand{\cbr}{chain-branching}
\newcommand{\gl}{global}
\newcommand{\ggl}{Global}
\newcommand{\me}{mechanism}
\newcommand{\mes}{mechanisms}
\newcommand{\sk}{skeletal}
\newcommand{\rd}{reduced}
\newcommand{\sts}{steady-state}
\newcommand{\nsts}{nonsteady-state}
\newcommand{\pe}{partial equilibrium}
\newcommand{\ff}{flow-field}
\newcommand{\il}{inner layer}
\newcommand{\ol}{oxidation layer}
\newcommand{\dam}{Damk{\"{o}}hler number}
\newcommand{\dams}{Damk{\"{o}}hler numbers}
\newcommand{\fcl}{fuel-consumption layer}
\newcommand{\neql}{nonequilibrium layer}
\newcommand{\rcl}{radical-consumption layer}
\newcommand{\dfl}{diffusion-flame layer}
\newcommand{\rnl}{radical-nonequilibrium layer}
\newcommand{\z}{{\xi}}
\newcommand{\zht}{{\xi}_{\rm {H_2}}}
\newcommand{\aig}{auto-ignition}
\newcommand{\aq}{a_{\rm {e}}}
\newcommand{\zst}{\xi_{\rm {st}}}
\newcommand{\xst}{x_{\rm {st}}}
\newcommand{\xp}{x_{\rm {p}}}
\newcommand{\zhtst}{{\xi}_{\rm {H_2,st}}}
\newcommand{\zp}{\xi_{\rm {p}}}
\newcommand{\zhtp}{{\xi}_{\rm {H_2,p}}}
\newcommand{\pchi}{\chi_{\rm {p}}}
\newcommand{\ipchi}{\chi_{\rm {p}}^{-1}}
\newcommand{\qpchi}{\chi_{\rm {p,q}}}
\newcommand{\stchi}{\chi_{\rm {st}}}
\newcommand{\nchi}{{\chi}^0}
\newcommand{\qstchi}{\chi_{\rm {st,q}}}
\newcommand{\inchi}{\chi_{\rm {I}}}
\newcommand{\zcp}{\xi_{\rm {p}}}
\newcommand{\zcr}{\xi_{\rm {r}}}
\newcommand{\tn}{T^0}
\newcommand{\tu}{T_{\rm {u}}}
\newcommand{\tc}{T_{\rm {c}}}
\newcommand{\yiu}{Y_{\rm {iu}}}
\newcommand{\tb}{T_{\rm {b}}}
\newcommand{\tst}{T_{\rm {st}}}
\newcommand{\tp}{T_{\rm {p}}}
\newcommand{\tfr}{T_{\rm {u}}}
\newcommand{\tfrr}{T_{\rm {ref}}}
\newcommand{\dtref}{{\Delta}T_{\rm {ref}}}
\newcommand{\tr}{T_{\rm {r}}}
\newcommand{\tl}{T_{\rm {l}}}
\newcommand{\mth}{methane}
\newcommand{\heptane}{{\it {n-}}heptane}
\newcommand{\decane}{{\it {n-}}decane}
\newcommand{\isooctane}{{\it {iso-}}octane}
\newcommand{\cisooctane}{{\it {iso-}}Octane}
\newcommand{\nbuta}{{\it {n-}}butanol}
\newcommand{\isobuta}{{\it {iso-}}butanol}
\newcommand{\isoprop}{{\it {iso-}}propanol}
\newcommand{\ethanol}{ethanol}
\newcommand{\ceth}{Ethanol}
\newcommand{\dme}{dimethyl ether}
\newcommand{\cdme}{Dimethyl ether}
\newcommand{\sheptane}{C$_7$H$_{16}$}
\newcommand{\sdecane}{C$_{10}$H$_{22}$}
\newcommand{\soctane}{C$_8$H$_{18}$}
\newcommand{\sdme}{CH$_3$OCH$_3$}
\newcommand{\seth}{C$_2$H$_5$OH}
\newcommand{\sisobuta}{C$_4$H$_8$OH}
\newcommand{\htt}{H$_2$}
\newcommand{\chf}{CH$_4$}
\newcommand{\ot}{O$_2$}
\newcommand{\nt}{N$_2$}
\newcommand{\hto}{H$_2$O}
\newcommand{\ssp}{$\,$}
\newcommand{\smo}{s$^{-1}$}

\newcommand{\tothe}[2]{#1$^{#2}$}
\newcommand{\eti}{\eta_{\rm {i}}}
\newcommand{\cci}{C_{\rm i}}
\newcommand{\echi}{H_{\rm i}}
\newcommand{\cpi}{c_{\rm p,i}}
\newcommand{\cpn}{c^0_{\rm p}}
\newcommand{\cm}{C_{\rm M}}
\newcommand{\cmn}{C_{\rm M}^0}
\newcommand{\cyi}{Y_{\rm {i}}}
\newcommand{\cyin}{Y_{\rm {I}}}
\newcommand{\cyott}{Y_{\rm {O_2,2}}}
\newcommand{\cyotr}{Y_{{\rm {O_2}},r}}
\newcommand{\cyotl}{Y_{{\rm {O_2}},l}}
\newcommand{\cyotst}{Y_{\rm {O_2,st}}}
\newcommand{\cyotln}{Y^0_{{\rm {O_2}},l}}
\newcommand{\cyot}{Y_{\rm {O_2}}}
\newcommand{\cyntot}{Y_{\rm {N_2O,2}}}
\newcommand{\cyf}{Y_{\rm {F}}}
\newcommand{\cyff}{Y_{\rm {F,1}}}
\newcommand{\cyhtf}{Y_{\rm {H_2,1}}}
\newcommand{\cxhtf}{X_{\rm {H_2,1}}}
\newcommand{\cyhtt}{Y_{\rm {H_2,2}}}
\newcommand{\cxhtt}{X_{\rm {H_2,2}}}
\newcommand{\cxi}{X_{\rm {i}}}
\newcommand{\sxi}{x_{\rm {i}}}
\newcommand{\cxf}{X_{\rm {F}}}
\newcommand{\cxfo}{X_{\rm {F,1}}}
\newcommand{\cxft}{X_{\rm {F,2}}}
\newcommand{\sxf}{x_{\rm {F}}}
\newcommand{\sxh}{x_{\rm {H}}}
\newcommand{\sxnto}{x_{\rm {N_2O}}}
\newcommand{\sxnt}{x_{\rm {N_2}}}
\newcommand{\syf}{y_{\rm {F}}}
\newcommand{\syh}{y_{\rm {H}}}
\newcommand{\cxot}{X_{\rm {O_2}}}
\newcommand{\cxnt}{X_{\rm {N_2}}}
\newcommand{\cxntt}{X_{\rm {N_2,2}}}
\newcommand{\cxott}{X_{\rm {O_2,2}}}
\newcommand{\cxotr}{X_{{\rm {O_2}},r}}
\newcommand{\cxotl}{X_{{\rm {O_2}},l}}
\newcommand{\cxotln}{X^0_{{\rm {O_2}},l}}
\newcommand{\cxntot}{X_{\rm {N_2O,2}}}
\newcommand{\cxntoo}{X_{\rm {N_2,1}}}
\newcommand{\cxnto}{X_{\rm {N_2O}}}
\newcommand{\sxot}{x_{\rm {O_2}}}
\newcommand{\echot}{H_{\rm {O_2}}}
\newcommand{\echnto}{H_{\rm {N_2O}}}
\newcommand{\echnt}{H_{\rm {N_2}}}
\newcommand{\echhto}{H_{\rm {H_2O}}}
\newcommand{\echcot}{H_{\rm {CO_2}}}
\newcommand{\echco}{H_{\rm {CO}}}
\newcommand{\echf}{H_{\rm {F}}}
\newcommand{\cxcht}{X_{\rm {CH_3}}}
\newcommand{\cxh}{X_{\rm {H}}}
\newcommand{\cxo}{X_{\rm {O}}}
\newcommand{\cxoh}{X_{\rm {OH}}}
\newcommand{\cxht}{X_{\rm {H_2}}}
\newcommand{\sxht}{x_{\rm {H_2}}}
\newcommand{\sxhtn}{(x_{\rm {H_2}})^0}
\newcommand{\cxhto}{X_{\rm {H_2O}}}
\newcommand{\sxhto}{x_{\rm {H_2O}}}
\newcommand{\cxhtop}{X_{\rm {H_2O,p}}}
\newcommand{\cxhtost}{X_{\rm {H_2O,st}}}
\newcommand{\cxcot}{X_{\rm {CO_2}}}
\newcommand{\sxcot}{x_{\rm {CO_2}}}
\newcommand{\cxcotp}{X_{\rm {CO_2,p}}}
\newcommand{\cxcotst}{X_{\rm {CO_2,st}}}
\newcommand{\cxcop}{X_{\rm {CO,p}}}
\newcommand{\cxntop}{X_{\rm {N_2O,p}}}
\newcommand{\cxco}{X_{\rm {CO}}}
\newcommand{\sxco}{x_{\rm {CO}}}
\newcommand{\cxotp}{X_{\rm {O_2,p}}}
\newcommand{\cyotp}{Y_{\rm {O_2,p}}}
\newcommand{\cxhtn}{X_{\rm {H_2}}^0}
\newcommand{\cxcon}{X_{\rm {CO}}^0}
\newcommand{\cxnton}{X_{\rm {N_2O}}^0}
\newcommand{\syi}{y_{\rm {i}}}
\newcommand{\sycht}{y_{\rm {CH_3}}}
\newcommand{\syht}{y_{\rm {H_2}}}
\newcommand{\syot}{y_{\rm {O_2}}}
\newcommand{\sznto}{{z}_{\rm {N_2O}}}
\newcommand{\szco}{{z}_{\rm {CO}}}
\newcommand{\szot}{{z}_{\rm {O_2}}}
\newcommand{\szn}{{z}^0}
\newcommand{\szcon}{{z}_{\rm {CO}}^0}
\newcommand{\szotn}{{z}_{\rm {O_2}}^0}
\newcommand{\szcoc}{{z}_{\rm {CO,c}}}
\newcommand{\szotc}{{z}_{\rm {O_2,c}}}
\newcommand{\sznton}{{z}_{\rm {N_2O}}^0}
\newcommand{\szht}{{z}_{\rm {H_2}}}
\newcommand{\szhtn}{{z}_{\rm {H_2}}^0}
\newcommand{\szhtc}{{z}_{\rm {H_2},c}}
\newcommand{\stn}{t^0}
\newcommand{\sztht}{\tilde{z}_{\rm {H_2}}}
\newcommand{\sztco}{\tilde{z}_{\rm {CO}}}
\newcommand{\sztcon}{\tilde{z}_{\rm {CO}}^0}
\newcommand{\tzeta}{\tilde{\zeta}}
\newcommand{\cdi}{D_{\rm {i}}}
\newcommand{\lei}{Le_{\rm {i}}}
\newcommand{\lef}{Le_{\rm {F}}}
\newcommand{\lecht}{Le_{\rm {CH_3}}}
\newcommand{\leot}{Le_{\rm {O_2}}}
\newcommand{\lento}{Le_{\rm {N_2O}}}
\newcommand{\leht}{Le_{\rm {H_2}}}
\newcommand{\leco}{Le_{\rm {CO}}}
\newcommand{\leh}{Le_{\rm {H}}}
\newcommand{\lehto}{Le_{\rm {H_2O}}}
\newcommand{\lecot}{Le_{\rm {CO_2}}}
\newcommand{\wbar}{\widehat{W}}
\newcommand{\wi}{W_{\rm {i}}}
\newcommand{\wf}{W_{\rm {F}}}
\newcommand{\win}{W_{\rm {I}}}
\newcommand{\wot}{W_{\rm {O_2}}}
\newcommand{\wnt}{W_{\rm {N_2}}}
\newcommand{\wnto}{W_{\rm {N_2O}}}
\newcommand{\wn}{w_{\rm {n}}}
\newcommand{\wk}{w_{\rm {k}}}
\newcommand{\omk}{{\omega}_{\rm {k}}}
\newcommand{\om}{\omega}
\newcommand{\kn}{k_{\rm {n}}}
\newcommand{\ckn}{K_{\rm {n}}}
\newcommand{\cqk}{Q_{\rm {k}}}
\newcommand{\nik}{\nu_{\rm {ik}}}
\newcommand{\nrho}{{\rho}^0}
\newcommand{\ptau}{\tau_{\rm {p}}}
\newcommand{\sttau}{\tau_{\rm {st}}}
\newcommand{\ntau}{\tau^0}
\newcommand{\p}{\rm {p}}
\newcommand{\qh}{q_{\rm {H}}}
\newcommand{\qf}{q_{\rm {F}}}
\newcommand{\qht}{q_{\rm {H_2}}}
\newcommand{\qco}{q_{\rm {CO}}}
\newcommand{\qnto}{q_{\rm {N_2O}}}
\newcommand{\sznls}{{\hat{z}}^0}
\newcommand{\inls}{\psi}
\newcommand{\szhtls}{{\hat{z}}_{\rm {H_2}}}
\newcommand{\szhtnls}{{\hat{z}}_{\rm {H_2}}^0}
\newcommand{\szcols}{{\hat{z}}_{\rm {CO}}}
\newcommand{\cxhtnls}{{\hat{X}}_{\rm {H_2}}^0}
\newcommand{\cxhtnss}{{\bar{X}}_{\rm {H_2}}^0}
\newcommand{\sznss}{\tilde{z}^0}
\newcommand{\szhtnss}{{\bar{z}}_{\rm {H_2}}^0}
\newcommand{\etac}{{\eta}_{\rm {c}}}
\newcommand{\neta}{{\eta}^0}
\newcommand{\aaa}{A}
\newcommand{\rrrt}{R}
\newcommand{\rrr}{R^0}
\newcommand{\eigenvalue}{\omega}
\newcommand{\cshat}{\overline{S}}
\newcommand{\cyhep}{Y_{\rm {hep}}}
\newcommand{\cyoct}{Y_{\rm {oct}}}
\newcommand{\cydme}{Y_{\rm {dme}}}
\newcommand{\cyeth}{Y_{\rm {eth}}}
\newcommand{\cyhepo}{Y_{\rm {hep,1}}}
\newcommand{\cyocto}{Y_{\rm {oct,1}}}
\newcommand{\cydmeo}{Y_{\rm {dme,1}}}
\newcommand{\cydmer}{Y_{{\rm {dme}},r}}
\newcommand{\cydmel}{Y_{{\rm {dme}},l}}
\newcommand{\cydmern}{Y^0_{{\rm {dme}},r}}
\newcommand{\cyetho}{Y_{\rm {eth,1}}}
\newcommand{\rrhep}{w_{\rm {hep}}}
\newcommand{\rroct}{w_{\rm {oct}}}
\newcommand{\rrdme}{w_{\rm {dme}}}
\newcommand{\rreth}{w_{\rm {eth}}}
\newcommand{\molhep}{W_{\rm {hep}}}
\newcommand{\moloct}{W_{\rm {oct}}}
\newcommand{\moldme}{W_{\rm {dme}}}
\newcommand{\moleth}{W_{\rm {eth}}}
\newcommand{\molot}{W_{\rm {O_2}}}
\newcommand{\diffhep}{D_{\rm {hep}}}
\newcommand{\diffoct}{D_{\rm {oct}}}
\newcommand{\diffeth}{D_{\rm {eth}}}
\newcommand{\diffdme}{D_{\rm {dme}}}
\newcommand{\diffot}{D_{\rm {O_2}}}
\newcommand{\cqhep}{Q_{\rm {hep}}}
\newcommand{\cqoct}{Q_{\rm {oct}}}
\newcommand{\cqdme}{Q_{\rm {dme}}}
\newcommand{\cqeth}{Q_{\rm {eth}}}
\newcommand{\zhep}{{\xi}_{\rm {hep}}}
\newcommand{\zoct}{{\xi}_{\rm {oct}}}
\newcommand{\zdme}{{\xi}_{\rm {dme}}}
\newcommand{\zeth}{{\xi}_{\rm {eth}}}
\newcommand{\zhepst}{{\xi}_{\rm {hep,st}}}
\newcommand{\zoctst}{{\xi}_{\rm {oct,st}}}
\newcommand{\zdmest}{{\xi}_{\rm {dme,st}}}
\newcommand{\zethst}{{\xi}_{\rm {eth,st}}}
\newcommand{\lehep}{Le_{\rm {hep}}}
\newcommand{\leoct}{Le_{\rm {oct}}}
\newcommand{\ledme}{Le_{\rm {dme}}}
\newcommand{\leeth}{Le_{\rm {eth}}}
\newcommand{\cxhep}{X_{\rm {hep}}}
\newcommand{\cxoct}{X_{\rm {oct}}}
\newcommand{\cxdme}{X_{\rm {dme}}}
\newcommand{\cxdmer}{X_{{\rm {dme}},r}}
\newcommand{\cxdmel}{X_{{\rm {dme}},l}}
\newcommand{\cxdmern}{X^0_{{\rm {dme}},r}}
\newcommand{\cxeth}{X_{\rm {eth}}}
\newcommand{\cxhepo}{X_{\rm {hep,1}}}
\newcommand{\cxocto}{X_{\rm {oct,1}}}
\newcommand{\cxdmeo}{X_{\rm {dme,1}}}
\newcommand{\cxetho}{X_{\rm {ethn,1}}}
\newcommand{\lphi}{{\phi}_l}
\newcommand{\rphi}{{\phi}_r}
\newcommand{\ssr}{a}
\newcommand{\qssr}{{\ssr}_{\mathrm {q}}}
\newcommand{\qssrpp}{{\ssr}_{\mathrm {q,pp}}}
\newcommand{\qssrnp}{{\ssr}_{\mathrm {q,np}}}
\newcommand{\sdistance}{l}
\newcommand{\cvelr}{V_{\mathrm {r}}}
\newcommand{\cvell}{V_{\mathrm {l}}}
\newcommand{\cvelf}{V_{\mathrm {1}}}
\newcommand{\cvelo}{V_{\mathrm {2}}}
\newcommand{\denr}{{\rho}_{\mathrm {r}}}
\newcommand{\denl}{{\rho}_{\mathrm {l}}}
\newcommand{\denf}{{\rho}_{\mathrm {1}}}
\newcommand{\deno}{{\rho}_{\mathrm {2}}}
\newcommand{\cyfr}{Y_{\mathrm {F,r}}}
\newcommand{\cyfl}{Y_{\mathrm {F,l}}}
\newcommand{\ffd}{flow-field}
\newcommand{\ppm}{partially premixed}
\newcommand{\cppm}{Partially premixed}
\newcommand{\od}{oxidizer-duct}
\newcommand{\ob}{oxidizer-boundary}
\newcommand{\tempf}{T_{1}}
\newcommand{\tempo}{T_{2}}
\newcommand{\cvlg}{V_{\rm {s}}}
\newcommand{\srf}{{a}_{1}}
\newcommand{\sro}{{a}_{2}}
\newcommand{\tg}{T_{\rm {s}}}
\newcommand{\jfs}{j_{\rm F,s}}
\newcommand{\jis}{j_{i,s}}
\newcommand{\cyis}{Y_{i,s}}
\newcommand{\brate}{{\dot{m}}}
\newcommand{\cyfg}{Y_{\rm F,s}}
\newcommand{\cxfg}{X_{\rm F,s}}
\newcommand{\tig}{T_{\rm {ig}}}


\newcommand{\one}{\rm {1}}
\newcommand{\onef}{\rm {1f}}
\newcommand{\oneb}{\rm {1b}}
\newcommand{\two}{\rm {2}}
\newcommand{\twof}{\rm {2f}}
\newcommand{\twob}{\rm {2b}}
\newcommand{\three}{\rm {3}}
\newcommand{\threef}{\rm {3f}}
\newcommand{\threeb}{\rm {3b}}
\newcommand{\four}{\rm {4}}
\newcommand{\fourf}{\rm {4f}}
\newcommand{\fourb}{\rm {4b}}
\newcommand{\five}{\rm {5}}
\newcommand{\nine}{\rm {6}}
\newcommand{\ninef}{\rm {6f}}
\newcommand{\nineb}{\rm {6b}}
\newcommand{\ten}{\rm {8}}
\newcommand{\eleven}{\rm {7}}
\newcommand{\elevenf}{\rm {7f}}
\newcommand{\elevenb}{\rm {7b}}
\newcommand{\twelvef}{\rm {12f}}
\newcommand{\twelveb}{\rm {12b}}
\newcommand{\thirteen}{\rm {9}}
\newcommand{\fourteen}{\rm {10}}
\newcommand{\kone}{k_{\one}}
\newcommand{\konef}{k_{\onef}}
\newcommand{\koneb}{k_{\oneb}}
\newcommand{\ktwo}{k_{\two}}
\newcommand{\ktwof}{k_{\twof}}
\newcommand{\ktwob}{k_{\twob}}
\newcommand{\kthree}{k_{\three}}
\newcommand{\kthreef}{k_{\threef}}
\newcommand{\kthreeb}{k_{\threeb}}
\newcommand{\kfour}{k_{\four}}
\newcommand{\kfourf}{k_{\fourf}}
\newcommand{\kfourb}{k_{\fourb}}
\newcommand{\kninef}{k_{\ninef}}
\newcommand{\knineb}{k_{\nineb}}
\newcommand{\kten}{k_{\ten}}
\newcommand{\ktwelvef}{k_{\twelvef}}
\newcommand{\ktwelveb}{k_{\twelveb}}
\newcommand{\kthirteen}{k_{\thirteen}}
\newcommand{\kfcm}{k_{\five}C_{\rm M}}
\newcommand{\kelevenf}{k_{\elevenf}}
\newcommand{\kelevenb}{k_{\elevenb}}
\newcommand{\kfourteen}{k_{\fourteen}}
\newcommand{\konen}{k_{\one}^0}
\newcommand{\konefn}{k_{\onef}^0}
\newcommand{\konebn}{k_{\oneb}^0}
\newcommand{\ktwon}{k_{\two}^0}
\newcommand{\ktwofn}{k_{\twof}^0}
\newcommand{\ktwobn}{k_{\twob}^0}
\newcommand{\kthreen}{k_{\three}^0}
\newcommand{\kthreefn}{k_{\threef}^0}
\newcommand{\kthreebn}{k_{\threeb}^0}
\newcommand{\kfourn}{k_{\four}^0}
\newcommand{\kfourfn}{k_{\fourf}^0}
\newcommand{\kfourbn}{k_{\fourb}^0}
\newcommand{\kninefn}{k_{\ninef}^0}
\newcommand{\kninebn}{k_{\nineb}^0}
\newcommand{\ktenn}{k_{\ten}^0}
\newcommand{\ktwelvefn}{k_{\twelvef}^0}
\newcommand{\ktwelvebn}{k_{\twelveb}^0}
\newcommand{\kthirteenn}{k_{\thirteen}^0}
\newcommand{\kfcmn}{k_{\five}^0C^0_{\rm M}}
\newcommand{\kelevenfn}{k_{\elevenf}^0}
\newcommand{\kelevenft}{k_{\elevenf}}
\newcommand{\kelevenbn}{k_{\elevenb}^0}
\newcommand{\kelevenbt}{k_{\elevenb}^0}
\newcommand{\kfourteenn}{k_{\fourteen}^0}
\newcommand{\done}{D_{\glI}}
\newcommand{\dtwo}{D_{\glII}}
\newcommand{\dfour}{D_{\glIV}}
\newcommand{\dfive}{D_{\glV}}
\newcommand{\ckone}{K_{{\one}}}
\newcommand{\cktwo}{K_{{\two}}}
\newcommand{\ckthree}{K_{{\three}}}
\newcommand{\ckfour}{K_{{\four}}}
\newcommand{\cknine}{K_{{\nine}}}
\newcommand{\ckeleven}{K_{{\eleven}}}
\newcommand{\ckonen}{K_{{\one}}^0}
\newcommand{\cktwon}{K_{{\two}}^0}
\newcommand{\ckthreen}{K_{{\three}}^0}
\newcommand{\ckfourn}{K_{{\four}}^0}
\newcommand{\ckninen}{K_{{\nine}}^0}
\newcommand{\wonef}{w_{{\onef}}}
\newcommand{\woneb}{w_{{\oneb}}}
\newcommand{\wtwof}{w_{{\twof}}}
\newcommand{\wtwob}{w_{{\twob}}}
\newcommand{\wthreef}{w_{{\threef}}}
\newcommand{\wthreeb}{w_{{\threeb}}}
\newcommand{\wfourf}{w_{{\fourf}}}
\newcommand{\wfourb}{w_{{\fourb}}}
\newcommand{\wfive}{w_{{\five}}}
\newcommand{\wninef}{w_{{\ninef}}}
\newcommand{\wnineb}{w_{{\nineb}}}
\newcommand{\wten}{w_{{\ten}}}
\newcommand{\welevenf}{w_{{\elevenf}}}
\newcommand{\welevenb}{w_{{\elevenb}}}
\newcommand{\wtwelvef}{w_{{\twelvef}}}
\newcommand{\wtwelveb}{w_{{\twelveb}}}
\newcommand{\wthirteen}{w_{{\thirteen}}}
\newcommand{\wfourteen}{w_{{\fourteen}}}
\newcommand{\glI}{\rm {I}}
\newcommand{\glII}{\rm {II}}
\newcommand{\glIII}{\rm {III}}
\newcommand{\glIV}{\rm {IV}}
\newcommand{\glV}{\rm {V}}
\newcommand{\hgo}{{\Delta}H_{\rm {G}}}
\newcommand{\hgoo}{{\Delta}H_{\rm {G1}}}
\newcommand{\hgot}{{\Delta}H_{\rm {G2}}}
\newcommand{\hgothree}{{\Delta}H_{\rm {G3}}}
\newcommand{\dthree}{D_{\glIII}}
\newcommand{\gthree}{G_{\glIII}}
\newcommand{\gtwo}{G_{\glII}}
\newcommand{\gkappa}{G_{\rm {b}}}
\newcommand{\galpha}{G_{\rm {a}}}
\newcommand{\mthree}{m_{\glIII}}
\newcommand{\mtwo}{m_{\glII}}
\newcommand{\mkappa}{m_{\rm {b}}}
\newcommand{\malpha}{m_{\rm {a}}}
\newcommand{\knn}{k^0_{\rm n}}
\newcommand{\cknn}{K^0_{\rm n}}
\newcommand{\ceonef}{E_{\onef}}
\newcommand{\ceoneb}{E_{\oneb}}
\newcommand{\cetwof}{E_{\twof}}
\newcommand{\cetwob}{E_{\twob}}
\newcommand{\cethreef}{E_{\threef}}
\newcommand{\cethreeb}{E_{\threeb}}
\newcommand{\ceninef}{E_{\ninef}}
\newcommand{\cenineb}{E_{\nineb}}
\newcommand{\alphaninef}{{\alpha}_{\ninef}}
\newcommand{\alphanineb}{{\alpha}_{\nineb}}
\newcommand{\alphafive}{{\alpha}_{\five}}


Experimental and computational investigations are carried out to elucidate the influence of {\ethanol} addition on {\heptane} {\aig} in counterflows. An axisymmetric stream of air, temperature gradually increased, is directed onto the surface of an evaporating pool of a liquid fuel. The air-stream temperature at {\aig} is measured at various strain rates, defined as the axial gradient of the axial component of the flow velocity at the stagnation plane, for {\heptane}, {\ethanol}, and various {\heptane}/{\ethanol} mixtures. Critical conditions for {\aig} are predicted employing the San Diego Mechanism for both fuels and the fuel mixtures, and the results are compared with the measurements.  Measurements and predictions show that low-temperature chemistry plays a significant role in promoting {\aig} of {\heptane} at low strain rates, but there is insufficient residence time at high strain rates for low-temperature chemistry to take place, so {\aig} is promoted by high-temperature chemistry. Experimental and computational results show that addition of {\ethanol} inhibits the low-temperature chemistry of {\heptane}. To identify the responsible elementary steps, computations are performed to identify those that dominate oxygen consumption and that contribute to the temperature rise in the reaction zone for {\heptane} and {\heptane}/{\ethanol} mixtures at low strain rates. For {\heptane} oxygen is consumed primarily by the low-temperature steps that result in ketohydroperoxide; the temperature rise is produced by subsequent low-temperature-chemistry steps. For the mixtures, a key step that consumes O$_2$ is O$_2$ + CH$_3$CHOH = HO$_2$ + CH$_3$CHO, and the heat release occurs through the classical high-temperature reaction mechanism. Thus, the inhibition of {\aig} that is observed to occur when {\ethanol} is added to {\heptane} arises from the competition for oxygen between this step and the low-temperature-chemistry addition of O$_2$ to the heptyl radical and to the radical arising from the subsequent isomerization, for {\heptane}.  

\end{abstract}
\vspace{10pt}
\parbox{1.0\textwidth}{\footnotesize {\em Keywords:} nonpremixed flows; autoignition; heptane; ethanol}
\rule{\textwidth}{0.5pt}
\vspace{10pt}

\end{@twocolumnfalse}] 

\clearpage

\section{Introduction} \addvspace{10pt}
\label{sec:introduction}
Commercial hydrocarbon fuels are often mixed with alcohols to enable clean and efficient combustion.  For this reason, a number of studies have been carried out on mixtures of alcohols with hydrocarbon fuels \cite{sarathy:2018:gasoline,gorbatenko:2019:butanol,goldsborough:2021:isoalcohol,mayhew:2021:fuelblending,yang:2013:nbutanol,welz:2013:biofuels,xu:2015:butanol,dalili:2020:isobutanol,cuoci:2021:surrogate,liang:2023:isobutanol,zhang:2013:heptane:buta,liu:2021:heptane}. These studies involved both practical fuel blends and ideal mixtures of components. Foremost among the combustion topics addressed was auto-ignition, since that is a key aspect relative to practical performance. The reader is referred to a useful review for further information concerning the extensive research performed prior to 2019, along with the motivation for the work \cite{sarathy:2018:gasoline}.\\

Starting from the early investigation of Tipper and Titchard \cite{tipper:1971:inhibition}, many studies have addressed the influence of alcohols on auto-ignition and combustion of hydrocarbon fuels \cite{tipper:1971:inhibition,cheng:2020:gasolineethanol,yang:2013:nbutanol,mayhew:2021:fuelblending,liang:2023:isobutanol,zhang:2013:heptane:buta}.  Tipper and Titchard \cite{tipper:1971:inhibition}  investigated  the effect of addition of large number of compounds on the cool-flame combustion of {\heptane} at around 533{\,}K in a static system. Addition of olefins and alcohols was found to inhibit the low-temperature chemistry of {\heptane}. Goldsborough et al.\ \cite{goldsborough:2021:isoalcohol} measured ignition delay times for {\aig} of mixtures of research-grade gasoline with {\isoprop} or  {\isobuta} at pressures of 20 and 40{\,}bar and temperatures from 700 to 1000{\,}K\@. A key finding of this investigation was that at low-temperature/NTC conditions (700-860 K) the iso-alcohols inhibit first-stage reactivity  of gasoline.  Similar results were obtained for ignition delay times measured in rapid compression machines for  mixtures of {\nbuta} and {\heptane} \cite{yang:2013:nbutanol} {\bf and in reflected shock waves \cite{zhang:2013:heptane:buta}.} Addition of {\nbuta} was found to increase the ignition delay times for values of pressure between 15{\,}bar and 30{\,}bar and temperatures between 650{\,}K and 830{\,}K indicating that {\nbuta} inhibits {\aig} of {\heptane} \cite{yang:2013:nbutanol}. Ignition delay times for {\aig} of mixtures of alcohols with jet-fuels and petroleum derived fuel were investigated at engine-relevant conditions in a pressure vessel \cite{mayhew:2021:fuelblending} at temperatures between 825 and 900{\,}K and pressures between 6 and 9{\,}MPa. The ignition delay times were found to increase with increasing addition of alcohol, thus confirming that alcohols inhibit low-temperature ignition of hydrocarbon fuels \cite{mayhew:2021:fuelblending}. {\bf  Other studies show that addition of ethanol to hydrocarbon fuels reduces emissions of oxides of nitrogen \cite{bogrek:2021:toluene:ethanol,lavadera:2021:heptaneblend} and influences formation of PAH and soot \cite{yan:2019:ethylene}.}  \\

Studies of {\aig} in rapid-compression machines and shock tubes are primarily concerned with  premixed systems and do not consider the influence of flow time on {\aig}. Recently, Liang et.\ al \cite{liang:2023:isobutanol} carried out an  experimental and computational investigation, employing the counterflow configuration, to elucidate the influence of {\isobuta} on critical conditions of {\aig} of {\decane} and {\heptane}. The temperature of the air stream at {\aig}, ${\tig}$, was measured at various values of the strain rate.  Kinetic modeling was carried out using the comprehensive CRECK chemical–kinetic mechanism.  Critical conditions of {\aig} were predicted and compared with the measurements. Low-temperature chemistry was found to play a significant role in promoting {\aig} of {\decane} and {\heptane}. Experimental data and numerical simulations showed that addition of even small amounts of {\isobuta} to {\decane} or {\heptane} increased the value of ${\tig}$ at low strain rates, indicating that {\isobuta} strongly inhibits the low-temperature chemistry of {\decane} and {\heptane}. Predicted flame structures showed that the peak values of mole fraction of ketohydroperoxide were significantly reduced when {\isobuta} was added to {\decane}, indicating that the kinetic pathway to low temperature ignition is blocked. This observation was confirmed by sensitivity analysis \cite{liang:2023:isobutanol}.\\

This previous study \cite{liang:2023:isobutanol} did not identify the steps specific to the kinetic model for alcohol combustion that are responsible for inhibiting low-temperature chemistry of  {\decane} and {\heptane}.  Here, an experimental and computational investigation is carried out to characterize the influence of addition of {\ethanol} (\seth) on {\aig} of {\heptane} (\sheptane), employing the counterflow configuration. Critical conditions for {\aig} are measured as a function of strain rate for various values of mixture ratios of the fuels. Computations are performed using the San Diego Mechanism \cite{sandiegomech} and the results are compared with measurements. A key goal of this investigation is to identify those kinetic steps in combustion of {\ethanol} that interfere with low temperature chemistry of {\heptane}, which has not been achieved in previous investigations.

\section{Experiments and Numerical Simulations} \addvspace{10pt}
\subsection{Experimental Apparatus and Procedures} 
\begin{figure}[h!]
\centerline{
 \includegraphics[width=192pt]{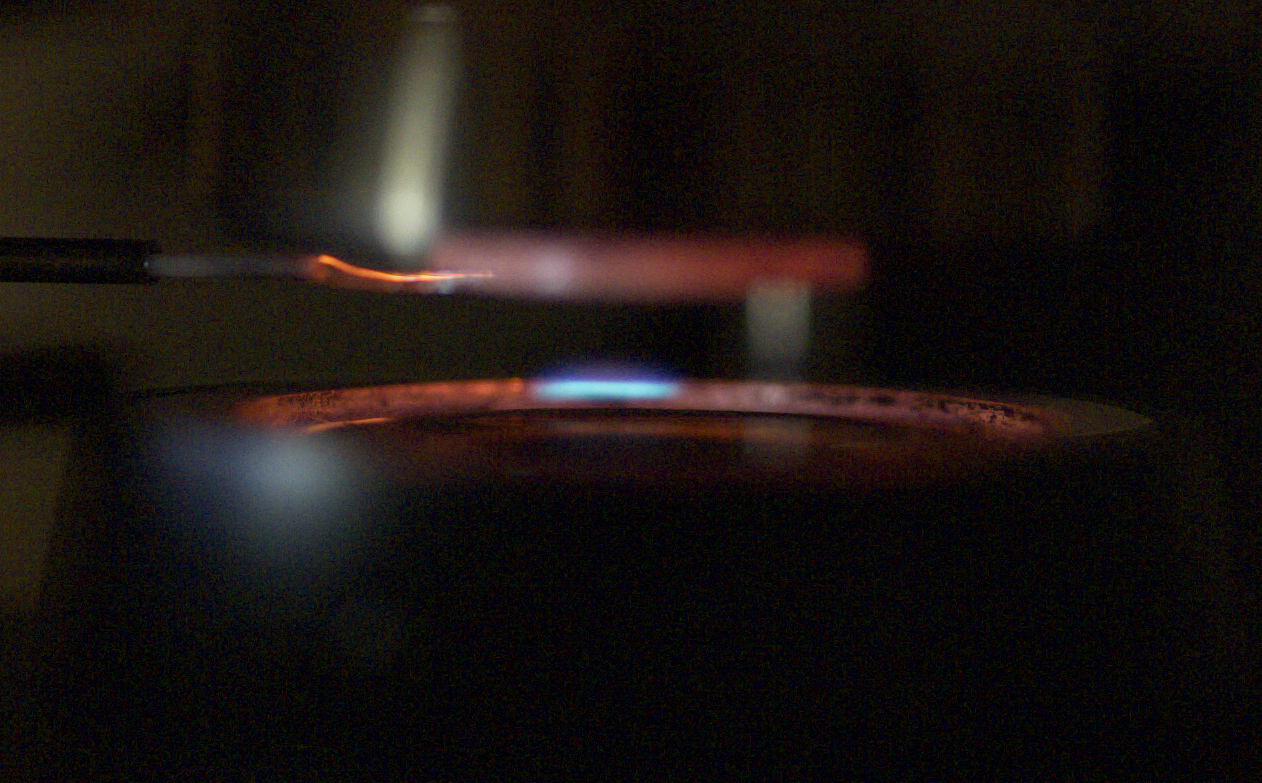}
}
\caption{High-speed photograph of the onset of {\aig} at strain rate ${\sro} = 100${\,}s$^{-1}$.  The fuel is 50{\,}{\%}{\,}{\heptane}/50{\,}{\%}{\,}{\heptane}.  The photo shows the rims of the fuel-cup and the oxidizer-duct, and the thermocouple employed to measure the {\aig} temperature of the {\os}.}
\label{fig:photo} 
\end{figure}

\addvspace{10pt}
\begin{figure}[h!]
\centerline{
 \includegraphics[width=192pt]{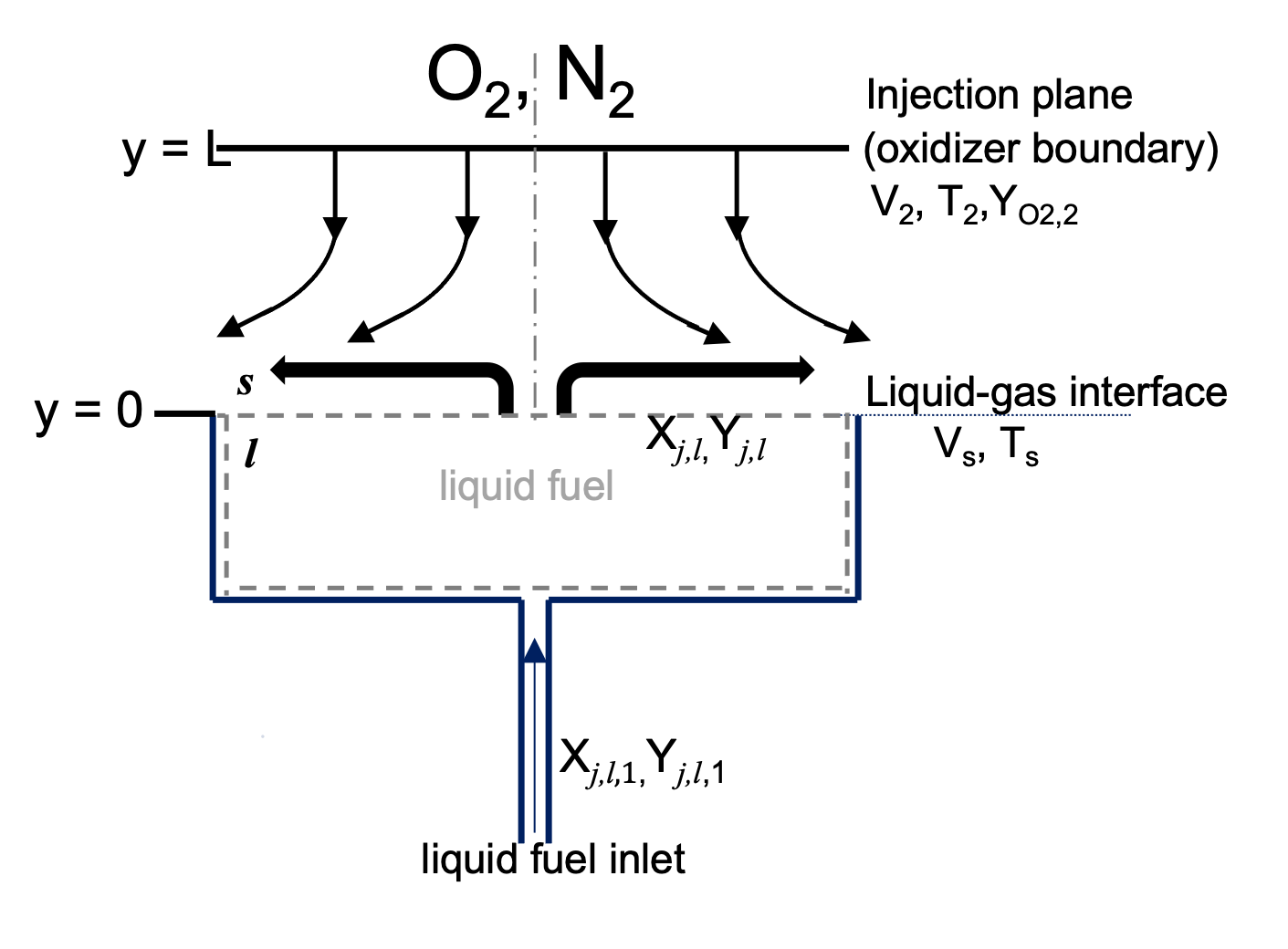}
}
\caption{Schematic illustration of the counterflow configuration. ${\cvelo}$ and ${\cvlg}$ are the velocities at the oxidizer boundary  and on  the gas side of the liquid-gas interface, respectively.  ${\tempo}$ and ${\tg}$ are the temperatures at the oxidizer boundary and the liquid-gas interface, respectively, and ${\cyott}$ is the mass fraction of oxygen at the oxidizer boundary.}
\label{fig:schematic} 
\end{figure}
Figure \ref{fig:schematic} is a schematic illustration of the “condensed-fuel” counterflow configuration employed in this experimental and computational study.  In this configuration, an axisymmetric flow of an {\os} made up of oxygen and nitrogen is directed over the surface of an evaporating pool of a liquid fuel in a fuel-cup.  It is injected from the {\od}, the exit of which is the oxidizer boundary\@. The origin is placed on the axis of symmetry at the surface of the liquid pool, and $y$ is  the axial co-ordinate and $r$ the radial co-ordinate and $y = 0$ represents the liquid-gas interface. The distance between the liquid-gas interface and the oxidizer boundary is $L$\@. At the oxidizer boundary $y = L$,  the magnitude of the injection velocity is ${\cvelo}$, the temperature  ${\tempo}$, the density ${\deno}$, and the mass  fraction of oxygen ${\cyott}$\@. Here, subscript $2$ represents conditions at the oxidizer boundary. The radial component of the flow velocity at the oxidizer boundary  is presumed to be equal to zero. The temperature at the liquid-gas interface is  ${\tg}$, and the mass averaged velocity on the gas side of the liquid-gas interface is ${\cvlg}$\@.  Here, subscripts $s$ and $l$, respectively, represent conditions on the gas-side  and the liquid-side of the liquid-gas interface. The quantities $X_{j,l}$ and $Y_{j,l}$ are, respectively, the mole-fraction and mass-fraction of the component $j$ in the liquid, and $X_{j,l,1}$ and $Y_{j,l,1}$ are, respectively, the mole-fraction and mass-fraction of the component $j$ in the liquid that is entering the fuel-cup of the counterflow burner. It has been shown previously \cite{seshadri:2008:ignition} that the radial component of the flow velocity at the  liquid-gas interface is small and can be presumed to be equal to zero. It has been shown that in the asymptotic limit of large Reynolds number the stagnation plane formed between the {\os} and the fuel vapors is close to the liquid-gas interface and a thin boundary layer is established there.  The inviscid flow outside the boundary layer is rotational. The local {\sr}, ${\sro}$, at the stagnation plane, is given by ${\sro} = 2{\cvelo}/L$ \cite{seshadri:1978:laminar,seshadri:2008:ignition}.  Figure \ref{fig:photo}
is a high-speed photograph of the onset of {\aig}.  When the critical condition of {\aig} is reached, a thin flame in the form of a disc first appears around the axis of symmetry above the liquid pool and subsequently rapidly covers the entire pool surface. \\

The counterflow burner is made up of two concentric tubes; an inner ceramic tube, and an outer quartz tube.  The ceramic tube has an inner diameter of $26{\,}$mm and an outer diameter of $28.6{\,}$mm\@. The {\os} flows through the inner tube and a curtain flow of nitrogen through the outer tube. The distance between the liquid-gas interface and the {\obd}, taking into consideration the thermal expansion of the {\od}, is $L = 10.5{\,}$mm\@. A  silicon carbide heating element, is placed inside the inner ceramic tube. The surface of the heating element can reach a temperature of 1900{\,}K\@.  All gaseous streams are controlled by computer regulated analog mass flow controllers. The velocity of the {\os} at the exit of the duct, ${\cvelo}$, is presumed to be equal to the ratio of the volumetric flow rate of the {\os} and the cross-section area of the duct.  The temperature of the oxidizer at the exit of the duct is measured using a Pt 10{\,}{\%} Rh/Pt 13{\,}{\%} Rh thermocouple with a wire diameter of 0.21{\,}mm and a bead diameter of 0.457{\,}mm. The thermocouple is held in place by a ceramic holder. The measured temperatures are corrected for radiative heat losses from the thermocouple bead using the Ranz and Marshall correlation for the Nusselt number for convective heat transfer from the gas to the spherical thermocouple bead \cite{heattransfer:book}. The repeatability of temperatures measured by the thermocouple is $\pm$ 5{\,}K\@.  Correction for radiative losses from the thermocouple bead are found to be approximately 20{\,}K, therefore the uncertainty in radiation correction is expected to be  $\pm$ 10{\,}K. \\

The procedure for measuring critical conditions of {\aig} is as follows.  First, the flow-field is established at a selected value of volumetric flow rate of the {\os}. Liquid fuel is introduced into the fuel-cup. The temperature of the {\os} is gradually increased in small increments, allowing sufficient time for the system to reach steady-state,  until {\aig} takes place. The velocity of the {\os} at the exit of the duct and the corresponding strain rate are calculated from the measured volumetric flow rate.  The temperature of air at {\aig}, ${\tig}$ is recorded as a function of the strain rate ${\sro}$. The experiment is repeated for different values of the strain rate. 
\subsection{Numerical Simulations} \addvspace{10pt}
The computations are performed using Cantera \cite{cantera:2023} C++ interface with modified boundary conditions for liquid-gas interface of liquid-pool \footnote{https://github.com/LJ1356/cantera.git}. 
Mix-average transport model is applied to obtained steady-state solutions. At the oxidizer boundary, the injection velocity ${\cvelo}$, the temperature, ${\tempo}$, and the value of ${\cyott}$ are specified. At the fuel side, equation (\ref{eq:bc:surface}) shows the boundary conditions for species conservation and energy conservation that are applied at the liquid-gas interface.
\begin{equation}
\label{eq:bc:surface}
\begin{array}{l}
  {\brate}{\cyis} + {\jis} = 0, \\
  {\brate}{Y_{j,s}} + j_{j,s} = {\brate}{Y_{j,l,1}}, \\
\left[{\lambda}\left(dT/dy\right)\right]_{\rm s} - {\brate}\sum_j Y_{j,l} h_{j,l}  = 0, \\
P_{v,j} X_{j,l} - p X_{j,s} = 0,
\end{array}
\end{equation}
and the constraint $\sum_j X_{j,l} -1 = 0$.  Here subscripts $i$ and $j$, respectively, refer to non-evaporating and evaporating species (specifically components of the liquid fuel),  ${\brate}$ is the mass evaporation rate, ${\cyis}$,  and  ${\jis}$ the mass fraction and diffusive flux of the non-evaporating species,  $Y_{j,s}$, $X_{j,s}$ and  $j_{j,s}$ the mass fraction, mole fraction and diffusive flux of the evaporating species on the gas side of the interface,  $\lambda$ is the thermal conductivity of the gas, and $h_{j,l}$, and $P_{\rm {v,j}}$, respectively, are the heat of vaporization and vapor pressure of component $j$ on the liquid-side of liquid-gas interface and $p$ the total pressure. The total mass flux of all species, $i$, on the gas-side of the liquid-gas interface comprises the diffusive flux, ${\jis}$, and the convective flux $\brate \cyis$.  The first expression in {\eqn}\ (\ref{eq:bc:surface}) imposes the condition that the total mass flux for all species, except for those of the evaporating fuel components, vanishes at the liquid-gas interface. The second expression of {\eqn}\ (\ref{eq:bc:surface}) imposes the constraint that the outgoing mass flux of each evaporating component in the liquid from the liquid-gas interface must be equal to the incoming mass flux, specifically the product of $\brate$ and the mass fraction of the species at liquid pool inlet, $Y_{j,l.1}$. The third expression in {\eqn}\ (\ref{eq:bc:surface}) is energy balance at the liquid-gas interface, and the fourth expression is Raoult's law relating the mole-fraction of the evaporating species on the gas side to the corresponding mole-fraction in the liquid.  \\ 

Kinetic modeling is carried out using the San Diego Mechanism \cite{sandiegomech}.  The computer program Canterra is used to compute the flame structure and critical conditions of auto-ignition. The fuels tested are {\heptane} (HPLC grade, purity $\geq$ 99{\,}{\%}), {\ethanol} and  mixtures with volumetric composition  of 20{\%}{\,}{\heptane}/80{\%}{\,}{\ethanol}, 50{\%}{\,}{\heptane}/50{\%}{\,}{\ethanol}, and  80{\%}{\,}{\heptane}/20{\%}{\,}{\ethanol}.  The oxidizer is air. {\bf The saturation vapor, $P_{\rm{v,j}}$ and the heat of vaporization $h_{j,l}$ in {\eqn}\ (\ref{eq:bc:surface}) for any species $j$ are evaluated using the expressions $\log_{10} P_{v,j} = A_{1, j} + B_j/T + C_j \times \log_{10} \left(T\right) + D_j \times T + F_j \times T^2$, and $h_{j,l} = A_{2, j} \left(1 - T/T_{j, cr}\right)^{N_j}$, where the value for the critical temperature $T_{j, cr}$ and the values for the empirical coefficients  $A_{1, j}, B_j, C_j, D_j, F_j, A_{2, j}$ and $N_j$ are obtained from \cite{yaws:2003}.} 




\section{Results and Discussion} 
\begin{figure}[h!]
\centerline{
   \includegraphics[width=192pt]{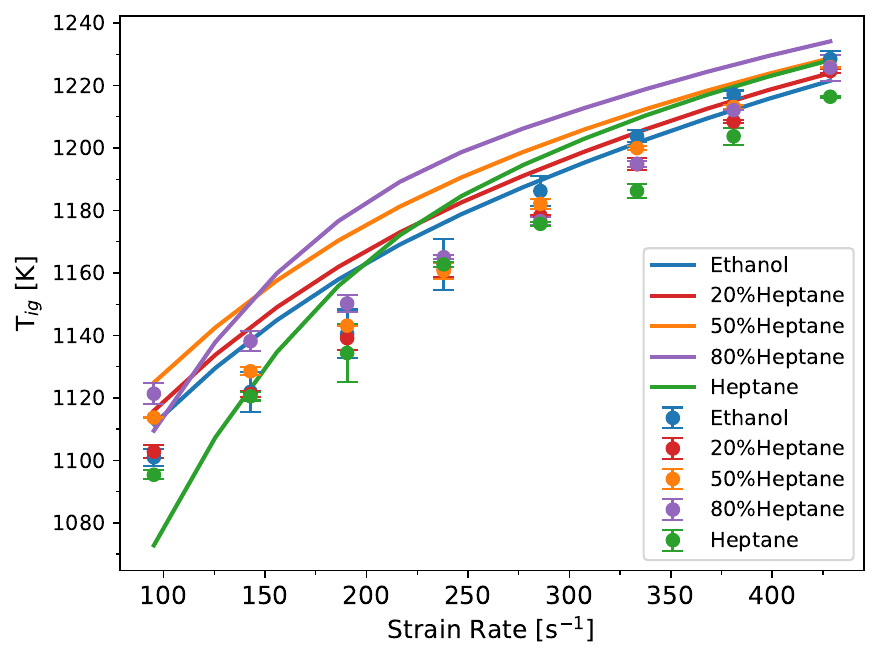}
}
\caption{The temperature of the air at {\aig}, ${\tig}$, of {\heptane}, {\ethanol} and mixtures of {\heptane} and {\ethanol} by volume as a function of strain-rate, ${\sro}$. The symbols represent experimental data and the lines are predictions. The uncertainty in experimental data is ${\pm}${\,}10{\,}K.}
\label{fig:expt:num} 
\end{figure}

\addvspace{10pt}
Figure \ref{fig:expt:num} 
shows the temperature of the air at {\aig}, ${\tig}$,  as a function of strain-rate, ${\sro}$, for {\heptane}, {\ethanol} and mixtures of these fuels.  In this figure, the symbols represent experimental data and the lines are predictions.  At low strain rates, around $\sro = 95${\,}s$^{-1}$, measurements show that {\heptane} is easiest to ignite because it has the lowest value of ${\tig}$ and the value of ${\tig}$ increases in the order, {\ethanol}, 20{\%}{\,}{\heptane}/80{\%}{\,}{\ethanol}, 50{\%}{\,}{\heptane}/50{\%}{\,}{\ethanol}, and  80{\%}{\,}{\heptane}/20{\%}{\ethanol}. It is noteworthy that at low strain rates, all mixtures have higher values of ${\tig}$ than the components of the mixture. At low strain rates computations show a similar trend where {\heptane} is easiest to ignite followed by {\ethanol} and  80{\%}{\,}{\heptane}/20{\%}{\ethanol} that have nearly the same value of ${\tig}$, while the mixtures 20{\%}{\,}{\heptane}/80{\%}{\,}{\ethanol},  and 50{\%}{\,}{\heptane}/50{\%}{\,}{\ethanol} have values of ${\tig}$ that are higher than those for {\heptane} and {\ethanol}. Moreover, experimental data and predictions show that  at low strain rates  addition of a small amount (20{\,}{\%}) of {\ethanol} increases ${\tig}$ by a  significant amount from that for {\heptane}, indicating that addition of {\ethanol} strongly inhibits the low-temperature chemistry of {\heptane}.  This behavior is similar to that observed in a previous investigation where {\isobuta} was found to inhibit low-temperature chemistry of {\heptane} and {\decane} \cite{liang:2023:isobutanol}. Figure \ref{fig:expt:num} shows that at high strain rates the measured value of ${\tig}$ for {\heptane} is the lowest and ${\tig}$ for  the mixtures are nearly the same as that for {\ethanol} and  the differences are well within experimental uncertainties. At high strain rates the predictions show that ${\tig}$ for {\ethanol} is the lowest followed by 20{\%}{\,}{\heptane}/80{\%}{\,}{\ethanol}, 50{\%}{\,}{\heptane}/50{\%}{\,}{\ethanol}, {\heptane} and  80{\%}{\,}{\heptane}/20{\%}{\ethanol}\@. Thus, the order of increase in values of ${\tig}$ in the experiment and predictions do not match at high strain rates. {\bf  In general, the quantitative agreement between the measurements and predictions are within experimental uncertainty. The deviations can also arise from uncertainties in the kinetic model for ethanol and requires further investigation}. \\

\begin{figure}[h!]
\centerline{
  \includegraphics[width=192pt]{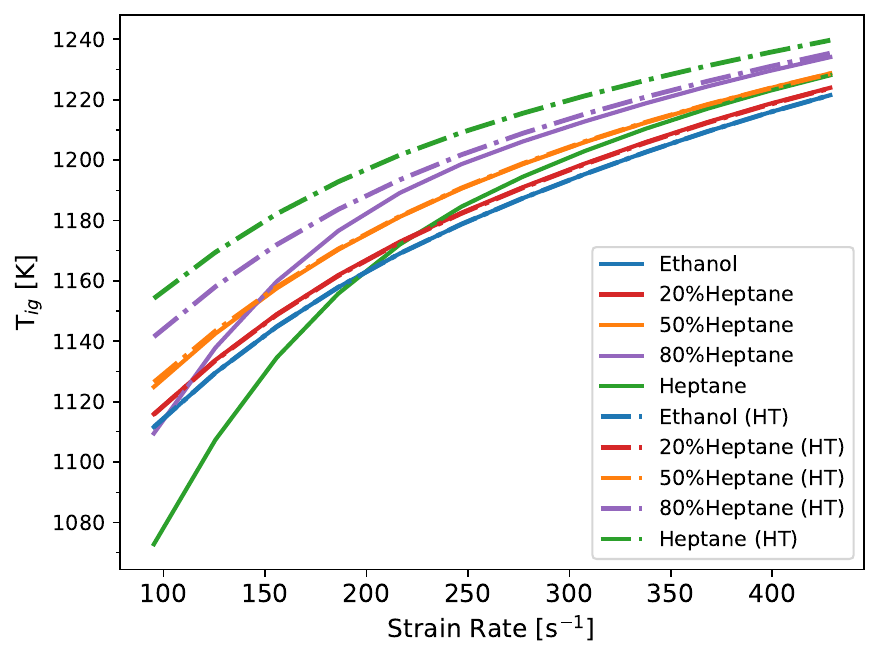}
}
\caption{The predicted temperature of the air at {\aig}, ${\tig}$,  of {\heptane}, {\ethanol} and mixtures of {\heptane}/{\ethanol} with percent volume ratios of  80/20, 50/50, and 20/80, as a function of strain-rate, ${\sro}$. The figure shows predictions with the complete kinetic mechanism and those with low-temperature chemistry removed (HT).}
\label{fig:num:ltht} 
\end{figure}

Following previous investigation where {\isobuta} was found to inhibit low-temperature chemistry of {\heptane} and {\decane} \cite{liang:2023:isobutanol}, computations were carried out with the complete mechanism and with the low-temperature reactions  of {\heptane} removed from the kinetic model and the results are shown in Fig.\  \ref{fig:num:ltht}.
For {\heptane}, at low strain rates, the value of ${\tig}$ calculated neglecting low-temperature chemistry is significantly larger than that predicted using complete  kinetic model. For the mixture with 80{\%}{\,}{\heptane}/20{\%}{\ethanol} ${\tig}$ calculated neglecting low-temperature chemistry is higher than that calculated using the complete model, but the differences are not as large as those for {\heptane}. Thus, some influence of low-temperature chemistry on {\aig} is still present in this mixture.  It is noteworthy that for the mixtures 20{\%}{\,}{\heptane}/80{\%}{\,}{\ethanol} and 50{\%}{\,}{\heptane}/50{\%}{\,}{\ethanol} the values of ${\tig}$ calculated with and without low-temperature chemistry are nearly the same, indicating that {\ethanol} has inhibited the low-temperature chemistry of {\heptane} for these mixtures. \\

\begin{figure}[h!]
\centerline{
  \includegraphics[width=192pt]{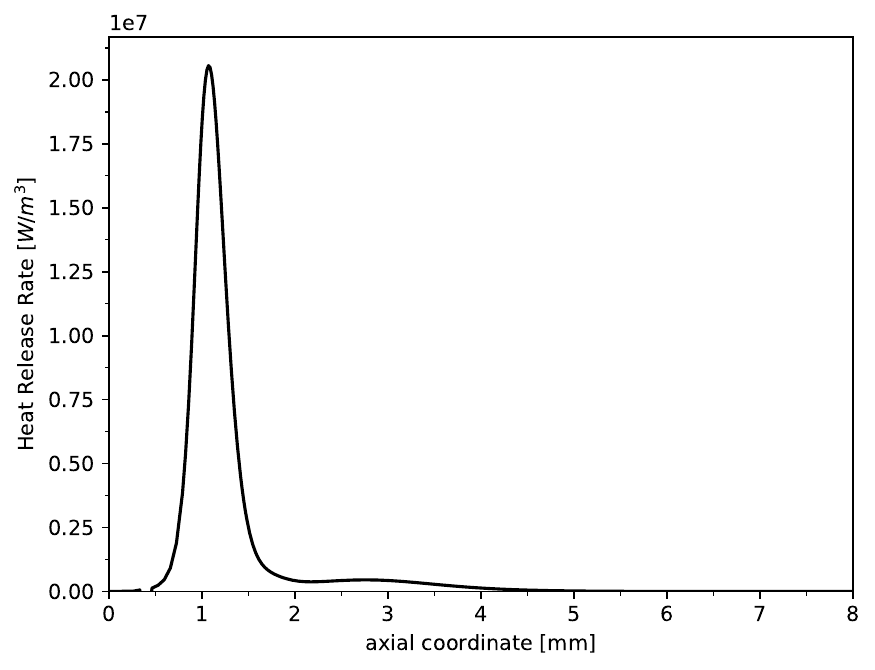}
}
\caption{Predicted profile of heat release for {\heptane}. Oxidizer temperature, $T_2 = 1000${\,}K, strain rate $\sro = 95${\,}s$^{-1}$.}
\label{fig:hr:heptane} 
\end{figure}

\begin{figure}[h!]
\centerline{
  \includegraphics[width=192pt]{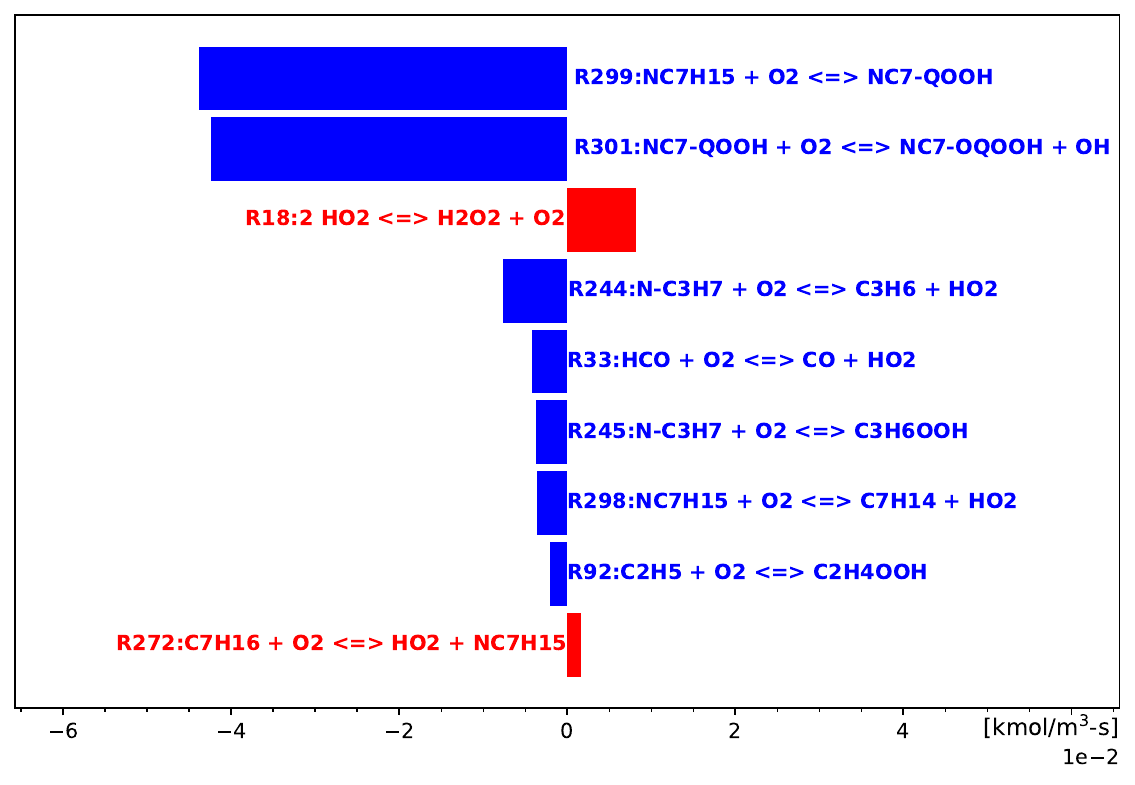}
}
\caption{Key elementary steps that consume oxygen for {\heptane} at the location of maximum heat release indicated in Fig.\ \ref{fig:hr:heptane} (y = 1.07{\,}mm). Oxidizer temperature, $T_2 = 1000${\,}K, strain rate ${\sro} = 95${\,}s$^{-1}$. Blue represents consumption and red formation.}
\label{fig:O2:heptane} 
\end{figure}

\begin{figure}[h!]
\centerline{
 \includegraphics[width=192pt]{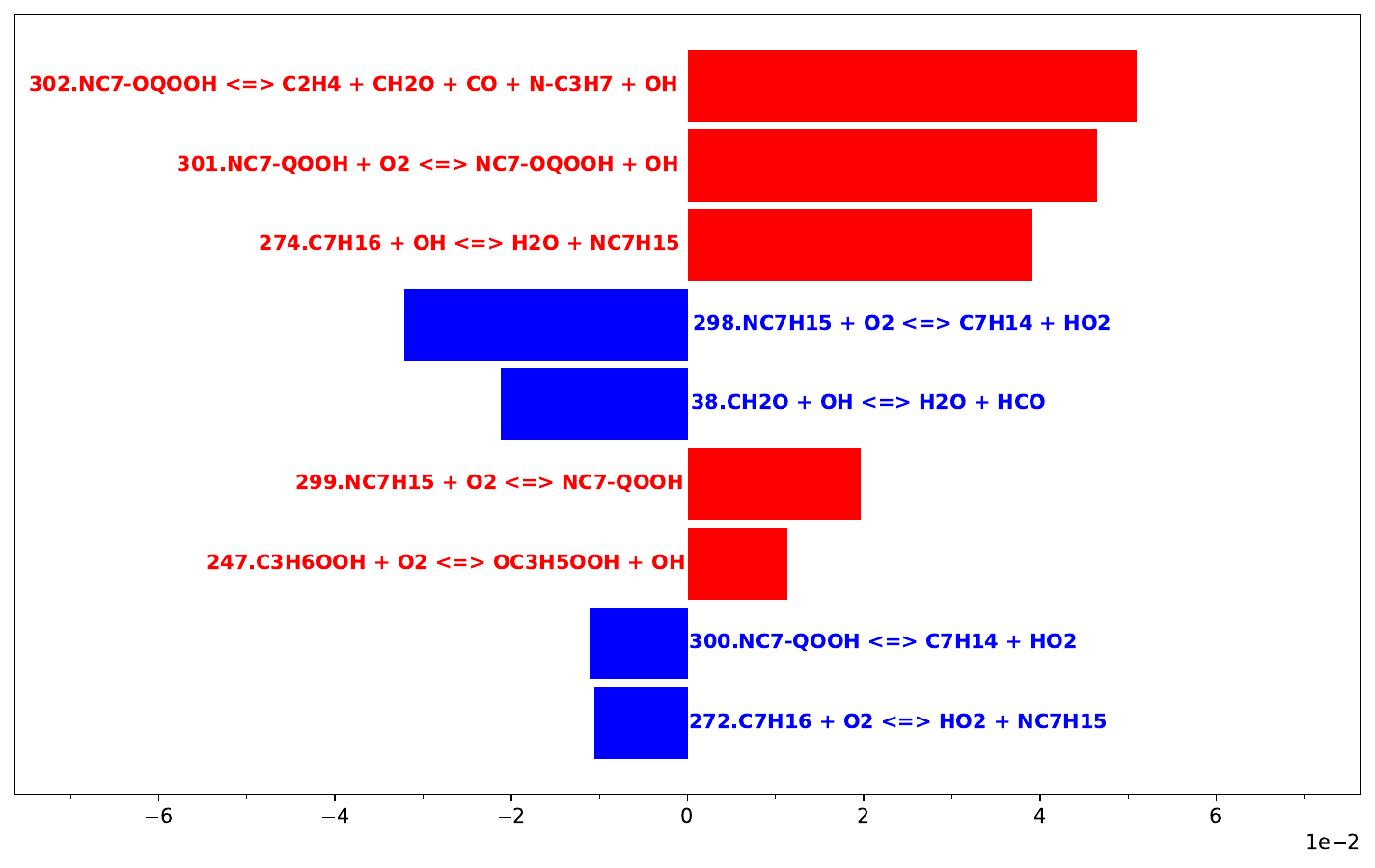}
}
\caption{Key elementary steps that contribute to the rise of temperature for {\heptane} at the location of maximum heat release indicated in Fig.\ \ref{fig:hr:heptane} (y = 1.07{\,}mm). Oxidizer temperature, $T_2 = 1000${\,}K, strain rate ${\sro} = 95${\,}s$^{-1}$. Blue represents reactions that decrease temperature and red reactions that increase temperature}
\label{fig:temp:heptane} 
\end{figure}

Figures \ref{fig:hr:heptane}, \ref{fig:O2:heptane}, and \ref{fig:temp:heptane} respectively, show predicted profiles of heat release, main elementary steps that consume oxygen, and the main elementary steps that contribute to the rise in temperature, close to {\aig} of {\heptane} at low strain rate, ${\sro} = 95${\,}s$^{-1}$. 
The liquid gas interface is at the axial location, $y = 0$ and the exit of the duct at $y = 10.5${\,}mm. Figure \ref{fig:hr:heptane} shows that the profile of heat release has two peaks, one around $y \approx 1{\,}$mm and the other around $y \approx 3{\,}$mm. The first peak, where low-temperature chemistry is expected to take place, is significantly higher than the second peak, where high temperature chemistry is expected to take place. Figures \ref{fig:O2:heptane} shows that O$_2$ is consumed primarily by the low-temperature reactions of, {\heptane} and Fig.\ \ref{fig:temp:heptane} shows that temperature rise is primarily due to low-temperature kinetic steps. Thus, for {\heptane} at low strain rates {\aig} is promoted by low-temperature chemistry. \\

{\bf Figure  \ref{fig:hr:ethanol}
\begin{figure}[h!]
\centerline{
  \includegraphics[width=192pt]{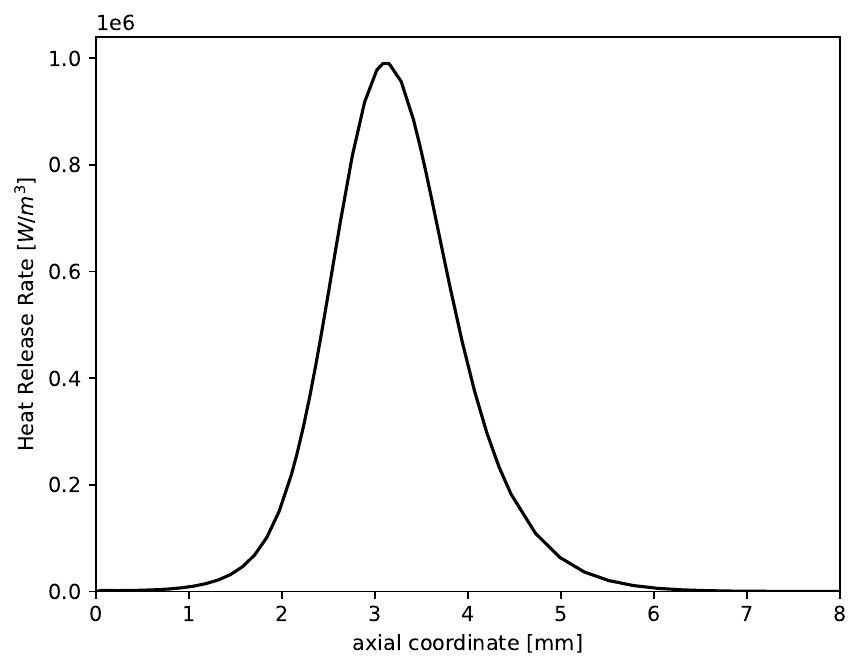}
}
\caption{Predicted profile of heat release for {\ethanol}. Oxidizer temperature, $T_2 = 1100${\,}K, strain rate $\sro = 95${\,}s$^{-1}$.}
\label{fig:hr:ethanol} 
\end{figure}
shows predicted profile of heat release close to {\aig} of {\ethanol} at low strain rate, ${\sro} = 95${\,}s$^{-1}$ and $T_2 = 1100{\,}K$. In contrast to the profile of heat-release for {\heptane} shown in Fig.\ \ref{fig:hr:heptane} the profile of heat release for {\ethanol} shows only one peak around 3{\,}mm.  This is consistent with the accepted point of view that unlike {\heptane},  {\ethanol} does not have separate low-temperature and high-temperature chemistry. }\\

\begin{figure}[h!]
\centering
\includegraphics[width=192pt]{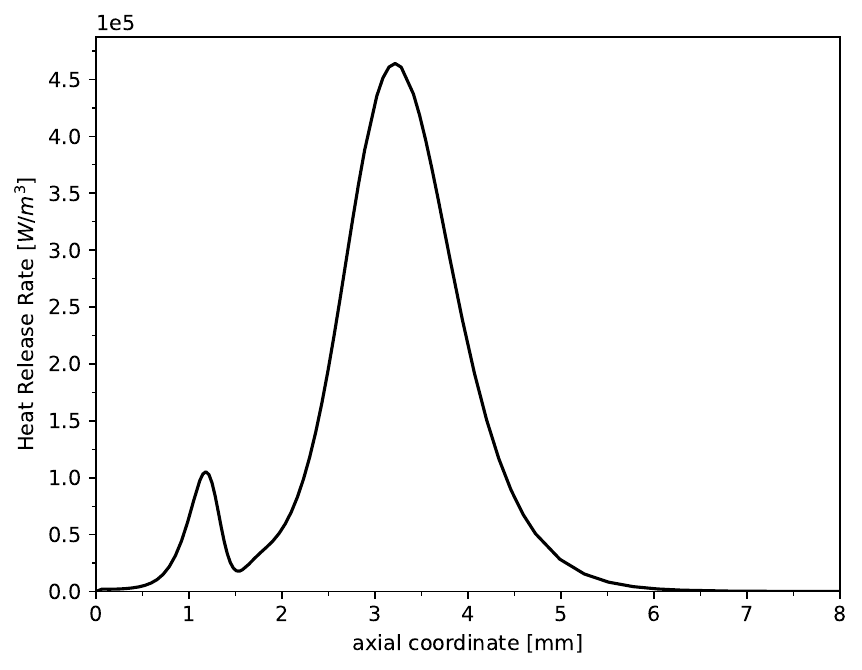}
\caption{Predicted profile of heat release for mixtures with volumetric composition  of 50{\%}{\,}{\heptane}/50{\%}{\,}{\ethanol}. Oxidizer temperature, $T_2 = 1100${\,}K, strain rate $\sro = 95${\,}s$^{-1}$.}
\label{fig:hr:50heptane} 
\end{figure}

\begin{figure}[h!]
\centering
\includegraphics[width=192pt]{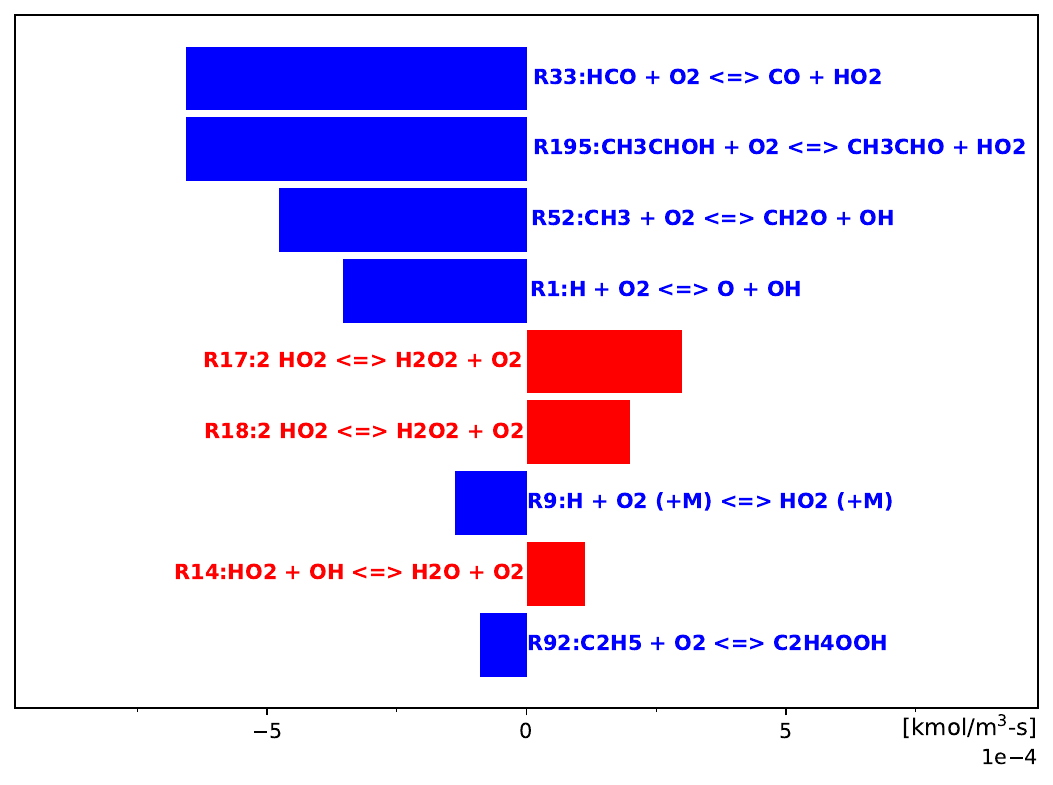}
\caption{Key elementary steps that consume oxygen for mixtures with volumetric composition  of 50{\%}{\,}{\heptane}/50{\%}{\,}{\ethanol} at the location of maximum heat release indicated in Fig.\ \ref{fig:hr:50heptane} (y = 3.22{\,}mm). Oxidizer temperature, $T_2 = 1100${\,}K, strain rate ${\sro} = 95${\,}s$^{-1}$. Blue represents consumption and red formation}
\label{fig:O2:50heptane} 
\end{figure}

\begin{figure}[h!]
\centering
\includegraphics[width=192pt]{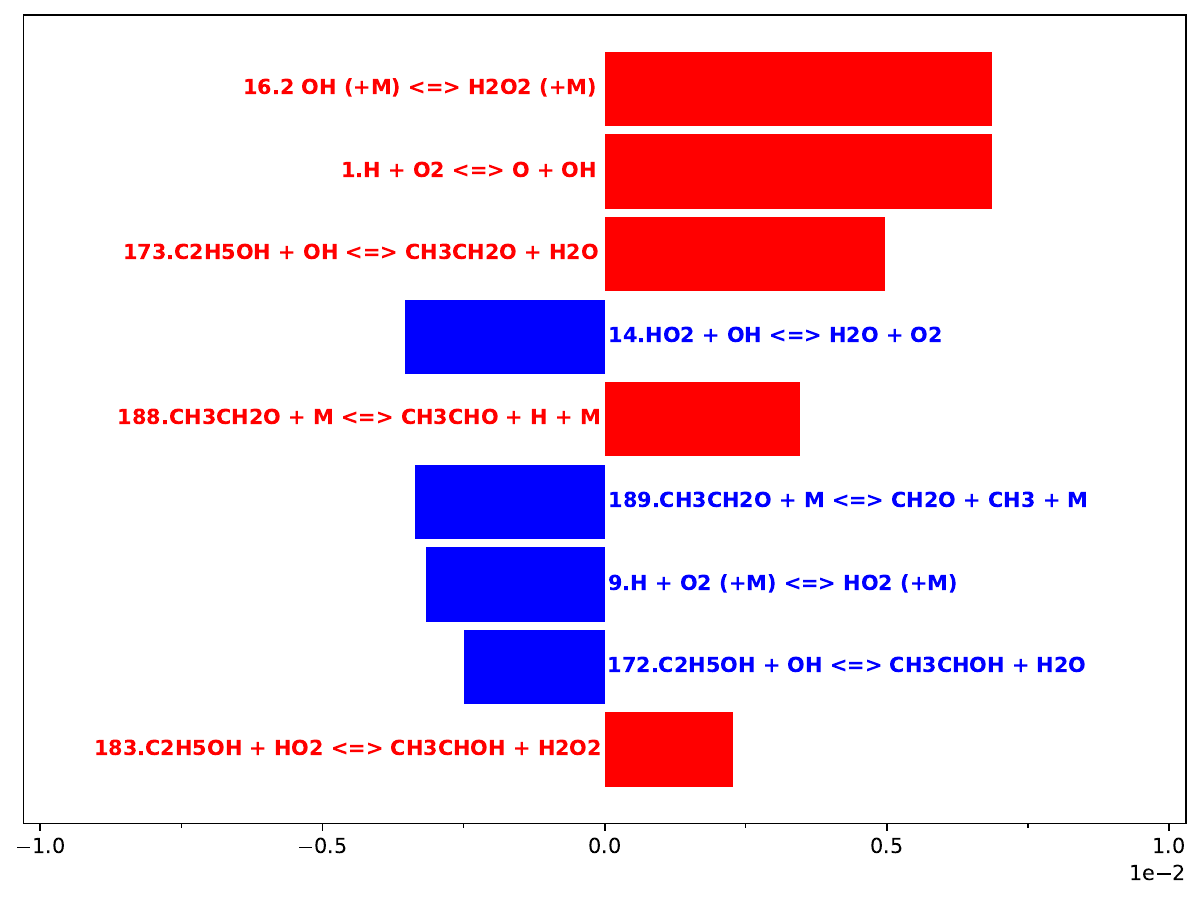}
\caption{Key elementary steps that contribute to the rise of temperature for mixtures with volumetric composition  of 50{\%}{\,}{\heptane}/50{\%}{\,}{\ethanol} at the location of maximum heat release indicated in Fig.\ \ref{fig:hr:50heptane} (y = 3.22{\,}mm). Blue represents reactions that decrease temperature and red reactions that increase temperature.}
\label{fig:temp:50heptane} 
\end{figure}

Figures \ref{fig:hr:50heptane}, \ref{fig:O2:50heptane}, and \ref{fig:temp:50heptane}, respectively, show the profile of heat release, main elementary steps that consume oxygen, and the main elementary steps that contribute to the rise in temperature for mixtures with volumetric composition  of 50{\%}{\,}{\heptane}/50{\%}{\,}{\ethanol}. The oxidizer temperature, $T_2 = 1100${\,}K and strain rate $\sro = 95${\,}s$^{-1}$.
The heat release profile in Fig.\  \ref{fig:hr:50heptane} shows two peaks at approximately the same locations as those in Fig.\ \ref{fig:hr:heptane}, however, in Fig.\  \ref{fig:hr:50heptane} the peak further away from the liquid-gas interface, where high-temperature reactions are expected to take place, is higher than the one closer to the liquid-gas interface. Figure \ref{fig:O2:50heptane} shows that a key step that consumes O$_2$ is O$_2$ + CH$_3$CHOH = HO$_2$ + CH$_3$CHO\@. Figure \ref{fig:temp:50heptane} shows that the temperature rise is primarily due to the reaction H + O$_2$ = OH + O and OH + OH = H$_2$O$_2$ + M which is different from that shown in Fig.\ \ref{fig:temp:heptane} where the temperature rise is from the low-temperature chemistry of {\heptane}. Thus, {\aig} for this mixture is primarily advanced by high-temperature reactions. \\

\begin{figure}[h!]
\centering
\includegraphics[width=192pt]{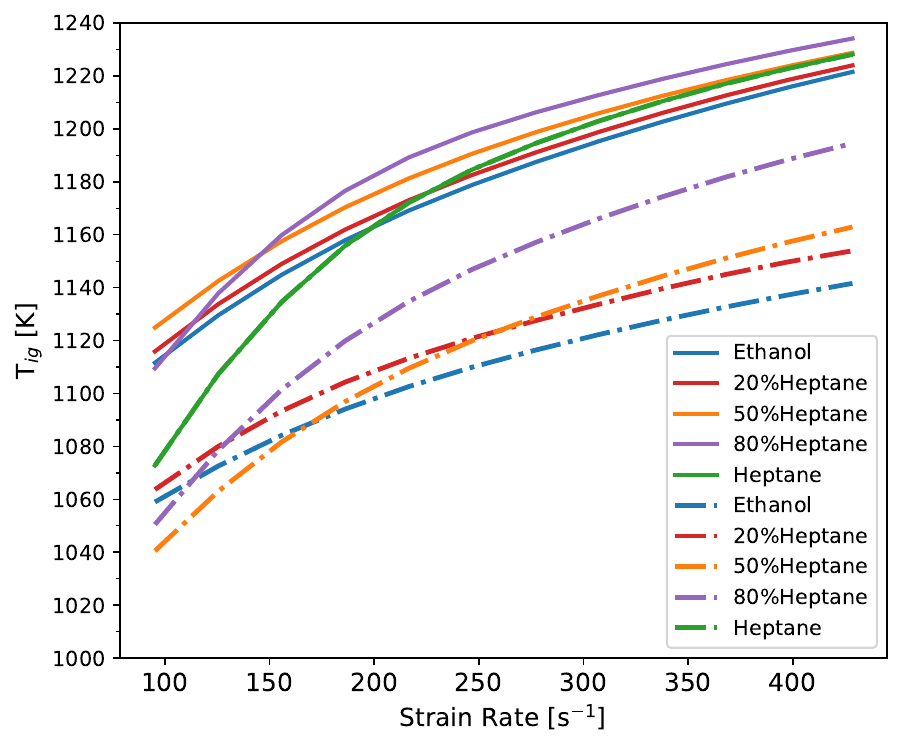}
\caption{The predicted temperature of the air at {\aig}, ${\tig}$,  of {\heptane}, {\ethanol} and mixtures of {\heptane}/{\ethanol} with percent volume ratios of 80/20, 50/50, and 20/80, as a function of strain-rate, ${\sro}$. The solid lines represent predictions with the complete kinetic mechanism, and the broken lines predictions with the step O$_2$ + CH$_3$CHOH = HO$_2$ + CH$_3$CHO removed.}
\label{fig:num:remove}
\end{figure}

To test if competition between kinetic steps that consume O$_2$ in the mechanism of {\heptane} and {\ethanol} are responsible for inhibition of {\aig} at low strain rates, computations were performed with the step  O$_2$ + CH$_3$CHOH = HO$_2$ + CH$_3$CHO removed from the kinetic model and the results are shown in Fig.\ \ref{fig:num:remove}. In this figure, the solid lines represent predictions with the complete kinetic mechanism and the broken line predictions with the step O$_2$ + CH$_3$CHOH = HO$_2$ + CH$_3$CHO removed.
As expected, Fig.\ \ref{fig:num:remove} shows that removal of this step does not change the critical conditions of {\aig} of {\heptane} but decreases the value of the {\aig} temperature, ${\tig}$, of {\ethanol} by a significant amount between 50 and 75{\,}K\@. 
It is noteworthy when this step  is removed, at low strain rates, the value of ${\tig}$ for the mixture 50{\,}{\%}{\,}{\heptane}/50{\,}{\%}{\,}{\ethanol} is lower than that for 20{\,}{\%}{\,}{\heptane}/80{\,}{\%}{\,}{\ethanol}, while this order is reversed at high strain rates. This behavior is a consequence of the fact that at low strain rates low-temperature chemistry is active, hence the mixture with the larger amount of {\heptane} has a lower value of ${\tig}$ while at high strain rates where low-temperature chemistry does not take place the order is reversed because the value of ${\tig}$ for {\heptane} is larger than that for {\ethanol} \cite{seshadri:2008:ignition,grana:2012:ignition}. Predictions including this step show that the value of ${\tig}$ for 50{\,}{\%}{\,}{\heptane}/50{\,}{\%}{\,}{\ethanol} is larger than that for 20{\,}{\%}{\,}{\heptane}/80{\,}{\%}{\,}{\ethanol} for all values of the strain rate. Thus, low-temperature chemistry of {\heptane} that was suppressed when {\ethanol} was added is restored when O$_2$ + CH$_3$CHOH = HO$_2$ + CH$_3$CHO is removed.   At high strain rates, Fig.\ \ref{fig:num:remove} shows that in the predictions with and without the step  O$_2$ + CH$_3$CHOH = HO$_2$ + CH$_3$CHO, the value of ${\tig}$ increases with increasing amounts of {\heptane} in the mixture because there is insufficient residence time for low-temperature chemistry to be active, therefore exclusion of this step does not have an influence on the critical conditions of {\aig}. \\

\begin{figure}[h!]
\centering
\includegraphics[width=192pt]{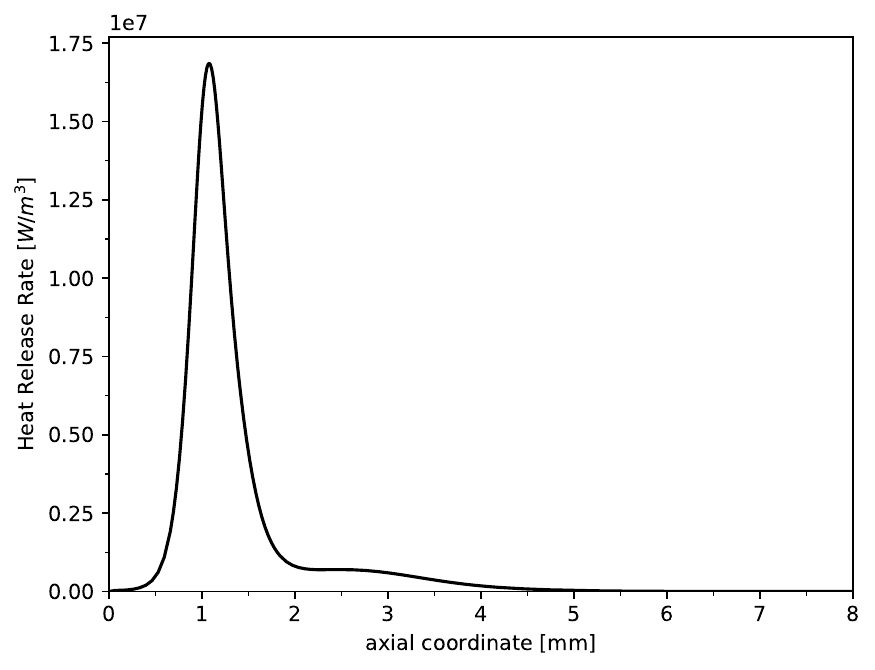}
\caption{Predicted profile of heat release for mixtures with volumetric composition  of 50{\%}{\,}{\heptane}/50{\%}{\,}{\ethanol} with step O$_2$ + CH$_3$CHOH = HO$_2$ + CH$_3$CHO removed and oxidizer temperature, $T_2 = 1000${\,}K, strain rate $\sro = 95${\,}s$^{-1}$.}
\label{fig:hr:remove} 
\end{figure}

\begin{figure}[h!]
\centering
\includegraphics[width=192pt]{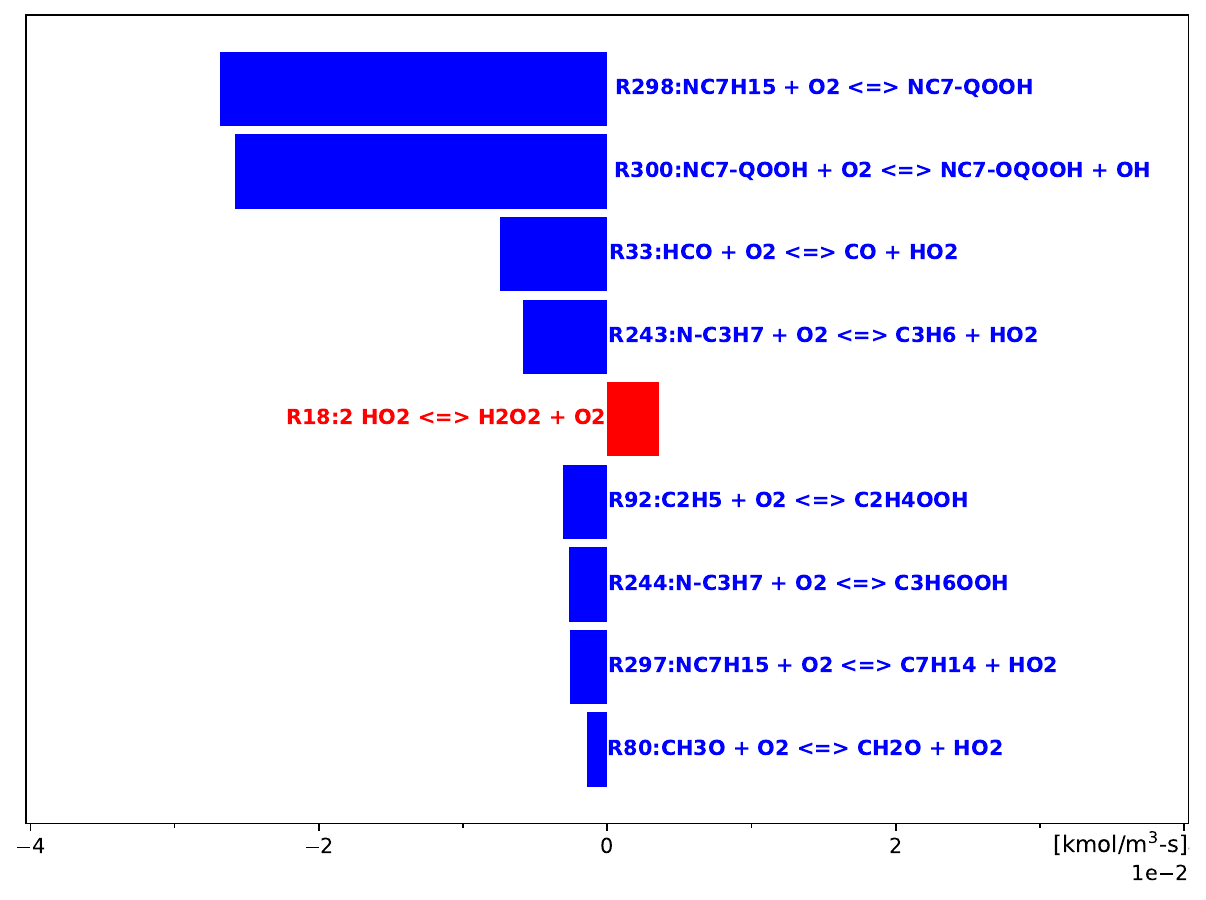}
\caption{Key elementary steps that consume oxygen for mixtures with volumetric composition  of 50{\%}{\,}{\heptane}/50{\%}{\,}{\ethanol} at the location of maximum heat release indicated in Fig.\ \ref{fig:hr:remove} (y = 1.1{\,}mm)  with step O$_2$ + CH$_3$CHOH = HO$_2$ + CH$_3$CHO removed and oxidizer temperature, $T_2 = 1000${\,}K, strain rate $\sro = 95${\,}s$^{-1}$. Blue represents consumption and red formation}
\label{fig:O2:remove}
\end{figure}

\begin{figure}[h!]
\centering
\includegraphics[width=192pt]{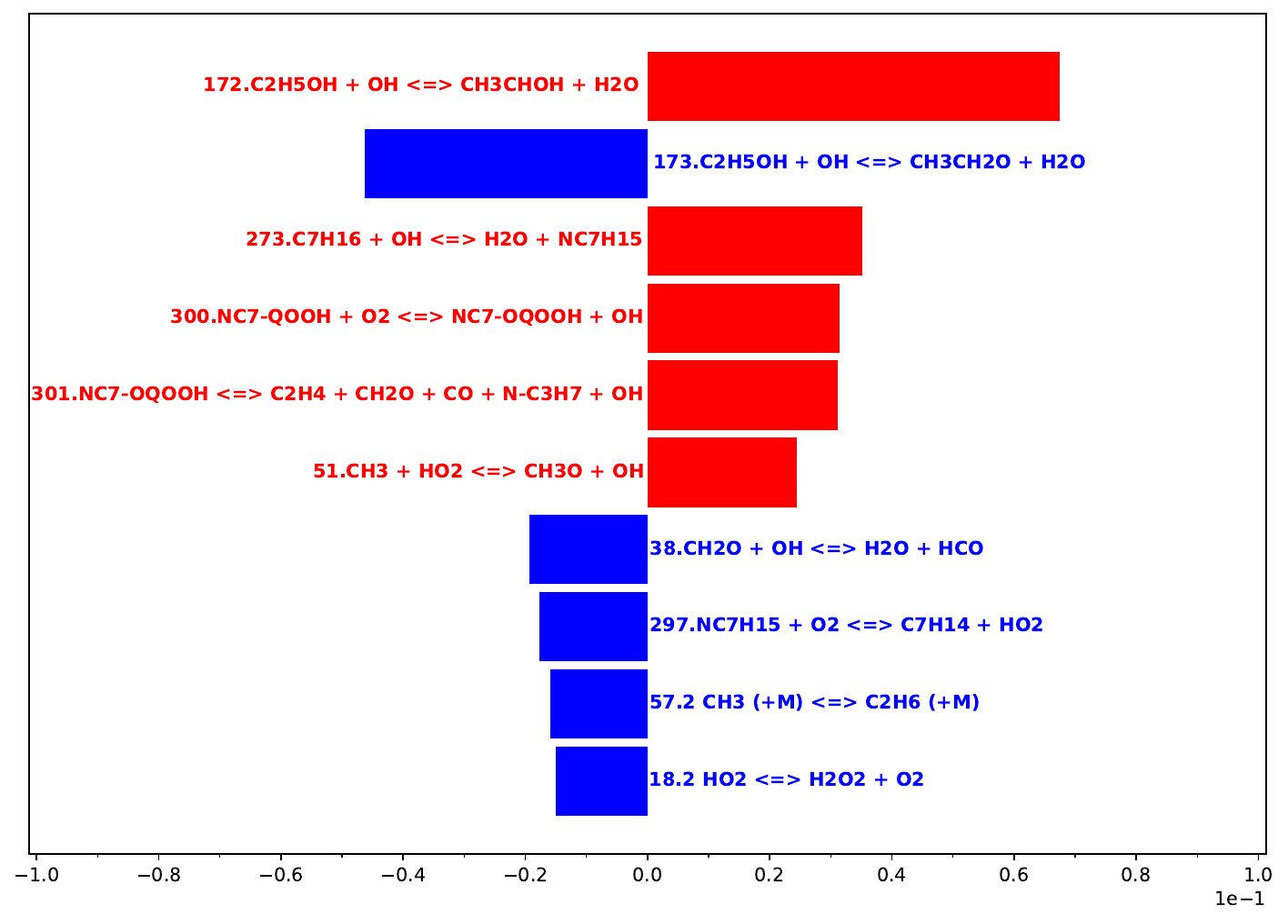}
\caption{Key elementary steps that contribute to the rise of temperature for mixtures with volumetric composition  of 50{\%}{\,}{\heptane}/50{\%}{\,}{\ethanol} at the location of maximum heat release indicated in Fig.\ \ref{fig:hr:remove} (y = 1.1{\,}mm) with step O$_2$ + CH$_3$CHOH = HO$_2$ + CH$_3$CHO removed and oxidizer temperature, $T_2 = 1000${\,}K, strain rate $\sro = 95${\,}s$^{-1}$. Blue represents reactions that decrease temperature and red reactions that increase temperature.}
\label{fig:temp:remove} 
\end{figure}

Figures \ref{fig:hr:remove}, \ref{fig:O2:remove}, and \ref{fig:temp:remove}, respectively, show profiles of heat release, main elementary steps that consume oxygen, and the main elementary steps that contribute to the rise in temperature predicted at conditions close to {\aig} for mixtures with volumetric composition  of 50{\%}{\,}{\heptane}/50{\%}{\,}{\ethanol} with  step O$_2$ + CH$_3$CHOH = HO$_2$ + CH$_3$CHO removed with $T_2 = 1000${\,}K, strain rate $\sro = 95${\,}s$^{-1}$. \\

Comparing the profile in of Fig.\ \ref{fig:hr:remove} with that in Fig.\ \ref{fig:hr:50heptane} shows that both figures show two peaks.  In the former, the peak closer to the liquid-gas boundary, where low-temperature chemistry is active, is more prominent while in the latter the peak further away from the liquid-gas boundary is more prominent.  This indicates that low-temperature chemistry of {\heptane} is restored when O$_2$ + CH$_3$CHOH = HO$_2$ + CH$_3$CHO is removed.  Figure \ref{fig:O2:50heptane} shows that O$_2$ is primarily consumed in the step O$_2$ + CH$_3$CHOH = HO$_2$ + CH$_3$CHO, while Fig.\ \ref{fig:O2:remove} shows that O$_2$ is consumed by the low-temperature steps of {\heptane}. Moreover,  Fig.\ \ref{fig:temp:50heptane} shows that the temperature rise in the reaction zone is primarily from high temperature chemistry, while Fig.\ \ref{fig:temp:remove} shows that the temperature increase is from the low-temperature reactions of {\heptane}.  These observations provide further confirmation that the step, O$_2$ + CH$_3$CHOH = HO$_2$ + CH$_3$CHO, competes with O$_2$ consumption by low-temperature reactions of {\heptane}. As a consequence, the low-temperature reactions of {\heptane} are suppressed when {\ethanol} is added. \\

{\bf Cheng et al. \cite{cheng:2020:gasolineethanol} studied {\aig} behavior of gasoline/ethanol blends in a rapid compression machine. In the low-temperature regime ethanol was found to retard first stage and main ignition delay times and suppress the rates and extents of low-temperature heat release.  Qualitatively this is similar to observations reported here where addition of ethanol not only increases the {\aig} temperature at low strain rates but also decreases the level of heat release in the region where low-temperature chemistry is expected to take place.}
\section{Concluding Remarks} \addvspace{10pt}

This work has identified the key mechanism through which ethanol addition inhibits the low-temperature {\aig} process of {\heptane}. Just as the heptyl radical exhibits an attractive site for addition of an oxygen molecule, so does the radical produced by H-atom abstraction from the ethanol site adjacent to the hydroxyl exhibit sufficient attraction for oxygen molecules to compete favorably with heptyl, yielding hydroperoxyl plus a stable molecule. By depriving heptyl and its isomerized oxygen-addition product (often denoted by QOOH in the literature) from a sufficient supply of oxygen molecules, the ethanol-generated radical turns off the low-temperature path in {\heptane}, thereby increasing its auto-ignition time. \\

This same mechanism is likely to prevail for higher alcohols, as well, thereby contribution to other perhaps unexpected experimental results. Future research involving these higher alcohols, as well as different  normal alkanes, would be worthwhile, to determine how generally relevant this type of new mechanism may be. Implications may be expected on {\aig} behaviors of developing environment-friendly new fuels designed to mitigate detrimental climate effect.


 \footnotesize
 \baselineskip 9pt


\bibliographystyle{pci}
\bibliography{seshadri,nsf,hydrocarbon,gasoline,aroproposal,highpressure,asymp,hydrogenaddition,biodieselproposal,petersnsf,coolflame,ethanol}


\newpage

\small
\baselineskip 10pt



\end{document}